\documentclass[twocolumn]{emulateapj}
\usepackage{psfig}

\newcommand{\pdif}[2]{\ensuremath{ \frac{\partial #1}{\partial #2}}}
\newcommand{\chia}{\chi^{\rm abs}}
\newcommand{\chis}{\chi^{\rm sc}}
\newcommand{\chit}{\chi^{\rm tot}}

\begin{document}
\title{A Radiation Transfer Solver for Athena using Short Characteristics}
\author{Shane W. Davis\altaffilmark{1}, James M. Stone\altaffilmark{2}, and Yan-Fei Jiang\altaffilmark{2}}
\altaffiltext{1}{Canadian Institute for Theoretical Astrophysics. Toronto, ON M5S3H4, Canada}
\altaffiltext{2}{Department of Astrophysical Sciences, Princeton University, Princeton, NJ 08544, USA}

\begin{abstract}

We describe the implementation of a module for the Athena
magnetohydrodynamics (MHD) code which solves the time-independent,
multi-frequency radiative transfer (RT) equation on multidimensional
Cartesian simulation domains, including scattering and non-LTE effects.
The module is based on well-known and well-tested algorithms developed
for modeling stellar atmospheres, including the method of short
characteristics to solve the RT equation, accelerated Lambda iteration
to handle scattering and non-LTE effects, and parallelization via
domain decomposition.  The module serves several purposes: it can be
used to generate spectra and images, to compute a variable Eddington
tensor (VET) for full radiation MHD simulations, and to calculate the
heating and cooling source terms in the MHD equations in flows where
radiation pressure is small compared with gas pressure.  For the latter
case, the module is combined with the standard MHD integrators using
operator-splitting: we describe this approach in detail, including
a new constraint on the time step for stability due to radiation
diffusion modes.  Implementation of the VET method for radiation
pressure dominated flows is described in a companion paper.  We present
results from a suite of test problems for both the RT solver itself,
and for dynamical problems that include radiative heating and cooling.
These tests demonstrate that the radiative transfer solution is accurate,
and confirm that the operator split method is stable, convergent, and
efficient for problems of interest.  We demonstrate there is no need to
adopt ad-hoc assumptions of questionable accuracy to solve RT problems in
concert with MHD: the computational cost for our general-purpose module
for simple (e.g. LTE grey) problems can be comparable to or less than
a single timestep of Athena's MHD integrators, and only few times more
expensive than that for more general (non-LTE) problems.

\end{abstract}


\section{Introduction}

Radiation is of fundamental importance for the thermodynamics of most
astrophysical systems.  It can be the dominant source of heating and
cooling of astrophysical plasmas.  Even in those systems where it plays
a minor role in energy transport, it is the dominant mechanism through
which we perceive and explore the universe.  Nevertheless, it has often
proven difficult to directly model the effects of radiation accurately
in modern multidimensional astrophysical (magneto)hydrodynamic (MHD)
codes due to both computational expense and conceptual complexity.

Most approaches to adding radiative transfer to dynamical
simulations are based on adopting restrictive assumptions or
approximations.  For example, often the flow is assumed to be
optically thin to radiation everywhere and for all time, or the
radiation field is assumed to originate in a small number of point
sources, with the diffuse emission from scattered or reradiated
photons ignored (such as in Cosmological reionization problems
e.g.\citealt{Abel:2002,Mellema:2006,Rijkhorst:2006,Whalen:2006,Reynolds:2009,Finlator:2009}). 

For problems in which the diffuse emission cannot be ignored, the dynamics
of the radiation field is often treated by solving the radiation
moment equations using ad hoc closure prescriptions to handle the
transition from optically thick to optically thin regimes, such as
flux-limited diffusion \citep[e.g.][ hereafter
  FLD]{Levermore:1981}. This includes applications such as accretion
flows, star formation, neutrino transport in supernovae, stellar
atmospheres and winds, cosmological reionization, and many others.
Indeed, there is a large and growing list of astrophysical MHD codes
that utilize FLD or a similar prescribed closure relation
\citep[including
  e.g.][]{Turner:2001,Bruenn:2006,Hayes:2006,Gonzalez:2007,Krumholz:2007a,Gittings:2008,Swesty:2009,Commercon:2011,van-der-Holst:2011,Zhang:2011}.

Numerical methods for directly solving radiative transfer (RT) have
been implemented
\citep[e.g.][]{Stone:1992,van-Noort:2002,Hayes:2003,Hubeny:2007}, but their
application to astrophysical problems has been somewhat limited,
especially in full 3D.  A notable exception is the progress made in
simulating the atmospheres of the Sun and other cool stars. In the
solar physics community, multidimensional MHD simulations of
convection with realistic RT have been performed for decades
with increasing sophistication. 
\citep[see e.g.][]{Nordlund:1982,Stein:1998,Vogler:2005,Heinemann:2007,Hayek:2010}

Encouraged by recent work modeling the departure of the radiation field
from local thermodynamic equilibrium (LTE) due to the presence of
electron scattering in three-dimensional MHD simulations \citep[see
  e.g.][]{Hayek:2010}, we have implemented a general-purpose RT solver
in Athena \citep{Stone:2008}, based on the methods widely used in the
stellar atmospheres community.  Athena is a general purpose
astrophysical MHD code, which is being actively developed and already
includes several modules for handling a variety of physical processes.
Effectively, we have combined Athena with a modern stellar atmospheres
code.  In fact, Athena already has a RT module that computes the
effects of ionization radiation from a single point source on the
surrounding gas \citep{Krumholz:2007}.  However, this module is not
well-suited for modeling the radiation from diffuse continuum
emission.

The addition of a RT solver to Athena enables three goals: (1) it can
be used as a diagnostic tool to compute self-consistently spectra and
images from time-dependent MHD simulations for direct comparison to
astronomical observations; (2) it allows us to compute a variable
Eddington tensor (VET) for the integration of the coupled MHD and
radiation moment equations (\citealt{Sekora:2010}; Jiang et al.,
submitted to ApJS, hereafter JSD12) for full radiation MHD simulations
in regimes where both energy and momentum transport by photons is
important; and (3) it allows us to compute the radiation source terms
in the energy equations and directly couple them to the MHD integrator
to compute the dynamics of flows where radiation pressure can be
ignored.

This paper focuses on describing our implementation of methods to
solve the RT equation, and the coupling of the solver with the MHD
integrator to compute the radiation source term in the energy equation.
The computation of the VET and solution of the radiation moment equations
is described in JSD12.  The plan of this work is as follows: In Section
\ref{equations} we summarize the equations that are solved.  In section
\ref{radtrans} we describe the detailed implementation of our solver and
the iterative methods used model deviations from LTE and handle certain
(e.g. periodic) boundary conditions.  In section \ref{interface} we
describe how we compute the radiation source terms in the energy equation
and incorporate them into the MHD integration.  In Section \ref{tests}
we present the results of several test problems not only to assess the
accuracy of the RT solver, but also to evaluate the performance of the
MHD integrator when the energy source terms are included.  We summarize
our results in Section \ref{summary}.

\section{MHD Equations with RT}
\label{equations}

In this work we solve the usual equations of compressible MHD, including
the source term in the energy equation to account for heating and cooling
due to radiation.  These source terms are computed directly from a formal
solution of the time-independent RT equation.  Thus, the basic equations
are continuity
\begin{equation}
\pdif{\rho}{t} + \mathbf{\nabla} \cdot \left(\rho \mathbf{v} \right) = 0 \label{eq:mass} ,
\end{equation}
momentum conservation
\begin{equation}
\pdif{\left(\rho\mathbf{v}\right)}{t} + \mathbf{\nabla} \cdot \left( \rho\mathbf{v} 
\mathbf{v} + {\sf T}\right) = 0,\label{eq:mom}
\end{equation}
the induction equation
\begin{equation}
\pdif{\mathbf{B}}{t} - \mathbf{\nabla} \times \left(\mathbf{v} \times \mathbf{B}\right) = 0,
\label{eq:induction}
\end{equation}
and energy conservation
\begin{equation}
\pdif{E}{t} + \mathbf{\nabla} \cdot \left(E \mathbf{v} + {\sf T} \cdot \mathbf{v}\right) = Q_{\rm rad}.
\label{eq:energy}
\end{equation}

In the above, $\rho$ is the gas density, $p$, $\mathbf{v}$ is the fluid
velocity, and $\mathbf{B}$ is the magnetic field.  The total stress
tensor $\sf T$ is defined as
\begin{equation}
  {\sf T} = (p + B^{2}/2){\sf I} - \mathbf{B}^{\rm T}\mathbf{B},
\label{eq:stresstensor}
\end{equation}
and $E$ is the total (fluid) energy
\begin{equation}
E=\frac{p}{\gamma-1}+\frac{1}{2}\rho v^2 + \frac{B^2}{2},
\end{equation}
where $p$ is the gas pressure and $\sf I$ is identity matrix.

The source term on the right hand side of equation (\ref{eq:energy}) is the
net gain or loss of energy due to radiative heating and cooling and is
given (for a static medium) by
\begin{equation}
 Q_{\rm rad} = 4 \pi \int^{\infty}_0 \chit_\nu \left(J_\nu - S_\nu \right) d \nu.
\label{eq:radsource}
\end{equation}
This is an integral over frequency $\nu$ of the difference between mean
intensity $J_\nu$ and the total source function $S_\nu$, weighted by
the total opacity\footnote{Note that $\chit_\nu$ has units of
  [cm$^{-1}$].  Throughout this work we will use $\chi$ for quantities
  with these dimensions and $\kappa = \chi / \rho$ for quantities with
  dimensions of [cm$^2$/g], but will refer to these interchangeably as
  opacities.}  We do not attempt to add the corresponding radiation
source term to the momentum equation.  This limits us to applications
in which radiation pressure is at most a modest fraction of gas
pressure.  An integrator for the coupled MHD and radiation moment
equations based on the one-dimensional algorithms discussed in
\citet{Sekora:2010} has been implemented in Athena and extended to
multidimensions by JSD12.  These more advanced techniques are needed
to handle the stiff source terms and modified dynamics in radiation
pressure dominated flows.

In order to compute the energy source term due to radiation, the MHD
equations must be
supplemented by the time-independent equation for RT
\begin{equation}
\hat{n} \cdot \nabla I_\nu = \chit_\nu \left(S_\nu - I_\nu\right),
\label{eq:radtrans}
\end{equation}
where $I_\nu$ is the specific intensity for an angle defined by the
unit vector $\hat{n}$.  In this work, we consider opacities due to
scattering $\chis_\nu$ and true absorption $\chia_\nu$, with
$\chit_\nu=\chia_\nu+\chis_\nu$.  It is convenient to define the
photon destruction probability $\epsilon_\nu = \chia_\nu/\chit_\nu$.
The source function is then given by
\begin{equation}
S_\nu = \epsilon_\nu B_\nu +(1-\epsilon_\nu) J_\nu,
\label{eq:sourcefunc}
\end{equation}
where $B_\nu$ is thermal source function.  The mean intensity $J_\nu$
is the ``zeroth'' moment, or average, of $I_\nu$ over solid angle
\begin{equation}
J_\nu  = \frac{1}{4\pi} \int I_\nu(\hat{n}) d\Omega. \label{eq:J}
\end{equation}

When absorption dominates
$\epsilon_\nu \rightarrow 1$ and $S_\nu \rightarrow B_\nu$, but when
scattering dominates $\epsilon_\nu \rightarrow 0$ and $S_\nu
\rightarrow J_\nu$.  Note that this expression assumes that scattering
is isotropic.  Although this is not strictly true for many scattering
processes (e.g. electron scattering), it will generally be a good
approximation for problems of interest.

In addition to $J_\nu$ we will also use
${\bf H}_\nu$ and ${\sf K}_\nu$, the first and second moments,
respectively.  Their components are given by
\begin{eqnarray}
H^i_\nu &=& \frac{1}{4\pi} \int I_\nu(\hat{n}) \mu_i d\Omega, \label{eq:H}\\
K^{ij}_\nu &=& \frac{1}{4\pi} \int I_\nu(\hat{n}) \mu_i \mu_j d\Omega, \label{eq:K}
\end{eqnarray}
where $d\Omega$ is the differential of solid angle, and $\mu_i \equiv
\hat{n} \cdot \hat{x}_i$.  These moments are related to the radiation
energy density $E_{\rm rad}$, radiation flux $\mathbf{F}_{\rm rad}$,
and radiation pressure ${\sf P}_{\rm rad}$ via the standard definitions
\begin{eqnarray}
E_{\rm rad}  &=& \frac{4\pi}{c} \int^{\infty}_0 J_\nu d \nu,\\
\mathbf{F}_{\rm rad} &=& 4\pi \int^{\infty}_0 \mathbf{H}_\nu d \nu,\\
{\sf P}_{\rm rad} &=& \frac{4\pi}{c} \int^{\infty}_0 {\sf K}_\nu d \nu.
\end{eqnarray}

Integration of equation (\ref{eq:radtrans}) over solid angle yields
\begin{equation}
-\mathbf{\nabla} \cdot \mathbf{F}_{\rm rad} = 4 \pi \int^{\infty}_0 \chit_\nu 
\left(J_\nu - S_\nu \right) d \nu.
\label{eq:rade}
\end{equation}
and provides an alternative (differential) form for the radiation
source term in equation (\ref{eq:energy}).  The differential form
tends to perform better in regions where optical depths across a
gridzone are large, while the integral form is preferable in regions
of low optical depth.  Hence, we will use both expressions, as
discussed in section \ref{interface}.

We have not been forced to make distinctions between the Eulerian and
comoving frame for radiation variables as we have dropped all velocity
dependent terms in equations (\ref{eq:radsource}),
(\ref{eq:radtrans}), and (\ref{eq:rade}).  We neglect these terms
because they are negligible for the tests considered in this paper.
However, we anticipate solving problems where the velocity dependent
terms may be important and can implement terms that are first order in
$v/c$ in our RT solver, where necessary.  For consistency with the VET
solver (JSD12), we will adopt the mix frame approach where $I_\nu$,
its moments, $\nu$, and $\hat{n}$ are Eulerian frame variables, while
opacities and emissivities are defined in the comoving frame.
Derivations of the mixed frame equations can be found in
\citet{Mihalas:1982}, \citet{Mihalas:1984}, \citet{Lowrie:1999}, and
\citet{Hubeny:2007}.

Since we neglect the time derivative of $I_\nu$ and terms that are first
order in $v/c$ in equation (\ref{eq:radtrans}), our method is only
formally reliable in the static diffusion and free streaming-limits.
Specifically, the timescale for fluid flow $t_{\rm f} \sim L/v$ across
a characteristic length scale $L$ in the simulation domain must be
longer than the time it takes for radiation to diffuse $t_{\rm dif}
\sim L^2 \chit/c$ or free-stream $t_{\rm fs} \sim L/c$ across the
domain \citep[see e.g.][]{Mihalas:1984}).  This is sufficient for the
test problems considered here and should be adequate for many of the
problems of primary interest to us.  When necessary, we can retain
terms first order in $v/c$ in equations (\ref{eq:radsource}) and
(\ref{eq:radtrans}) and the code will be formally accurate in the
dynamic diffusion limit ($t_{\rm f} \lesssim t_{\rm dif}$) as well.

Throughout this work $B_\nu$ is assumed to correspond to the Planck
function and is a function only of $\nu$ and gas temperature $T$.  We
assume an ideal equation of state with $p=\rho R T$ and gas thermal
energy density $E_{\rm gas}=p/(\gamma-1)$.  Here $R$ is the gas
constant and $\gamma$ is the adiabatic index.  The adiabatic sound
speed is $a=\sqrt{\gamma p/\rho}$.

The methods for solving the MHD equations without the radiation source
term are described in detail in previous publications
\citep{Gardiner:2008,Stone:2008,Stone:2009} and are unchanged by the
solution of radiation transfer.  The computation of RT is described in
Section \ref{radtrans} and the interface of the RT solver and MHD
integrator is described in Section \ref{interface}.  The sequence for
a single timestep can summarized as follows:

1) Using the hydrodynamic variables (typically $T$ and $\rho$) from
the previous timestep as inputs, we compute $\chit_\nu$, $\epsilon_\nu$, and
$B_\nu$, or each frequency in each grid zone.

2) We solve Equation (\ref{eq:radtrans}) using the methods described
in Section \ref{radtrans}, yielding $S_\nu$ and $J_\nu$ everywhere in
the domain.

3) Using $S_\nu$ and $J_\nu$ (or $H_\nu$), we compute the radiation
source term $Q_{\rm rad}$ and update equation (\ref{eq:energy}) as
described in Section \ref{interface}.

4) We advance the MHD variables using the standard Athena integrators.

\section{Solution of Radiation Transfer}
\label{radtrans}

An extensive literature on the solution of RT for astrophysical
problems in multidimensions exists and there are numerous monographs
and review articles on the topic
\citep[e.g.][]{Mihalas:1984,Castor:2004,Carlsson:2008}. With this
literature to draw from, we have largely adopted a strategy of
implementing existing, well-developed algorithms.  Since there are
many different methods with different strengths and weaknesses, the
major challenge is finding a method which best suits our particular
needs.  Our most salient constraints include:

1) The method needs to be amenable to domain decomposition since this
is the primary algorithm for parallelizing the solution of the MHD
equations in Athena.

2) The method must be able to handle the explicit dependence of the
source term on $J_\nu$ in equation (\ref{eq:sourcefunc}) for problems
in which scattering is important (i.e. we must be able solve non-LTE
problems).

3) The method needs to be able to handle (shearing) periodic boundary
conditions.

4) The method must be robust and capable of handling
discontinuities in temperature and density which arise when
shocks are present in the flow.

5) Ideally, the method should be efficient enough that for simple
problems (e.g. LTE with grey or mean opacities), neither the memory
constraints nor the total computational time is dominated by the
solution of RT.

With these considerations in mind, we have implemented a
short-characteristics based solver
\citep{Mihalas:1978,Olson:1987,Kunasz:1988}.  In this method the
specific intensity is discretized on a set of rays at each cell center
in the simulation domain.  Equation (\ref{eq:radtrans}) is integrated
along each ray using initial intensities interpolated from neighboring
grid zones. Since only neighboring grid zones are used for this
integration, the total computational cost (per iteration) scales
linearly with the number of gridzones in the domain.  This is also
simple to parallelize with domain decomposition as only information
from cells on the faces of the neighboring sub-domains need to be
passed.

This is in contrast to a long characteristics method
\citep[e.g.][]{Feautrier:1964}, which would integrate the RT equation
along each ray through all gridzones intersected by the ray until the
edges of the simulation domain are reached.  Such a method is
generally more computationally expensive since computation of the
specific intensity in each gridzone typically requires integrating
equation (\ref{eq:radtrans}) through $\sim N^{1/3}$ gridzones (where
$N$ is the total number of gridzone in the domain). It is also more
cumbersome to use with domain decomposition \citep[see,
however,][]{Heinemann:2006} since it may require the passing of
larger blocks of data, including information from non-neighboring
subdomains.

Although short characteristic methods are computationally more
expedient, they suffer from greater numerical diffusion due to the
interpolation that is required to compute the intensity in neighboring
gridzones \citep{Kunasz:1988}. For problems where a few gridzones (or
point sources) dominate the total emissivity, a short characteristics
solver may require very high angular resolution to accurately resolve
the radiation field far from the dominant source.  If the angular
resolution is too low, anomalous structure (e.g. spokes) in the
heating and cooling rates will emanate from the dominant sources
\citep[see e.g.][]{Finlator:2009}.  (In this case the numerical
diffusion introduced by interpolation can be beneficial.)  Instead,
the emission from point sources is better handled by suitably designed
long characteristics methods (\citealt{Abel:2002,Krumholz:2007},
although see also \citealt{Rijkhorst:2006}).  For the applications of
interest in this work (e.g. accretion flows), the diffuse radiation
field dominates.  Moreover, even when point sources are present, the
diffuse radiation field due to scattering or re-emission (e.g. HII
regions) cannot generally be ignored, and therefore we anticipate such
problems may be accommodated in the future by a hybrid scheme which
uses short characteristics for the diffuse emission, and long
characteristics for bright point sources.

Non-LTE problems are handled via iteration.  For each time step the
formal solution of the whole domain is repeated, updating $J_\nu$ and
$S_\nu$ during each iteration, until some formal convergence criterion
is met.  As discussed below, we implement an accelerated (or
approximate) lambda iteration (hereafter ALI) algorithm based on the
Gauss-Seidel method of \citet{Trujillo-Bueno:1995} (hereafter TF95).
The TF95 method is efficient for solving non-LTE problems because it
significantly increases the convergence rate without significantly
increasing the computational cost (or memory footprint) per iteration.

Iteration is also used in LTE problems to handle boundary conditions
at the interface of subdomains and for physical periodic boundary
conditions at domain edges.  On each iteration the incoming intensity
from the neighboring subdomain is fixed from the previous iteration
(or timestep for the first iteration).  The outgoing intensity, which
corresponds to the incoming intensity in the neighboring subdomain, is
then updated and the formal solution is iterated to convergence.  For
LTE problems, this is not the most efficient method for handling the
subdomain boundaries \citep[see e.g.][]{Heinemann:2006}.  For the
moment, we are primarily interested in non-LTE problems where
iteration is required regardless.  We generally find fairly rapid
convergence (requiring only a few iterations) for most of our LTE test
problems when iterations is used, so this is not a
significant limitation.

In many respects our short-characteristics RT solver is similar to
those of \citet{van-Noort:2002} and \citet{Hayek:2010} in that both
implement ALI to handle deviations from LTE and both utilize domain
decomposition for parallelization.  \citet{Hayek:2010} used their code
to solve the RT equation including scattering, in MHD simulations of
stellar atmospheres on three-dimensional Cartesian grids.  Hence, the
effectiveness of several key aspects of our module have already been
demonstrated in a sophisticated MHD code and applied to realistic
astrophysical applications.

\subsection{Frequency Discretization}

The scheme we have implemented allows for the computation of frequency
dependent, grey, or monochromatic RT.  Radiation variables (moments
and specific intensities) and radiative properties of the fluid such
as the opacities, thermal source function, and photon destruction
parameter are tabulated on a grid of $n_{\rm f} \ge 1$ discrete
frequencies or frequency groups.  For flexibility, the functional form
of opacities and emissivities can be specified via user-defined
functions.  In general, the computational cost and memory footprint of
problems scale linearly with $n_{\rm f}$.

These frequency bins can simply be discrete frequencies when RT is
used to generate diagnostic outputs such as images and spectra.  Group
mean opacities and emissivities \citep[e.g.][]{Mihalas:1984,
  Skartlien:2000} and corresponding quadrature weights must be
specified when the RT solver is used to compute the radiation source
terms or VET.  In the simplest case, $n_f=1$ and an appropriate
frequency integrated mean opacity is specified.

Unless otherwise noted, we will drop subscripts denoting the frequency
dependence of radiation variables and only describe the monochromatic
problem hereafter.  For the problems under present consideration,
there is no explicit coupling of the specific intensity at different
frequencies so the frequency dependent calculation is a trivial
generalization of the monochromatic problem.

\subsection{Angular Discretization}

We discretize the specific intensity on both angular and spatial
grids.  For one-dimensional problems, the discretization is chosen so
that polar angles correspond to the abscissas for Gaussian quadrature.
In multidimensions, discretization of the angles proceeds according to
the algorithm described in Appendix B of \citet{Bruls:1999}, which is
based on the principles of type A quadrature described in
\citet{Carlson:1963}.

This method attempts to distribute the rays as evenly as possible over
the unit sphere, subject to the constraint that each octant of the
unit sphere is discretized identically.  Hence the angle
discretization is invariant for 90$^\circ$ rotations about the
coordinate axes.  This is desirable because Athena is designed to be a
general purpose code, and there is often no preferred direction with
which to align the angular grid (as in some atmosphere calculations).
Without this constraint, the result would generally depend on the
orientation of boundary and initial conditions relative to the
coordinate axes.

The user specifies the number of polar angles $n_\mu$, and the code
generates an array of $n_{\rm a}$ rays covering the unit sphere.  In
one dimension, this corresponds to $n_{\rm a} = n_\mu$ rays because of
axisymmetry.  For multidimensional domains $n_{\rm a} = n_\mu (n_\mu
+2)$ rays.  However, in two-dimension only half of these are unique
due to the implied invariance of physical quantities in the third
dimension and $n_{\rm a} = n_\mu (n_\mu +2)/2$. 

Setting $n_\mu = 2$ in a one-dimensional calculation is analogous to
invoking the two-stream approximation, in which the radiation field of
each hemisphere is approximated by transfer along a single ray.  This
assumption is commonly used to derive analytic solutions, and allows
the ratio of $H/J$ to vary but keeps the ratio of $K/J$ fixed at 1/3,
consistent with the Eddington approximation.  In two (three)
dimensional calculations, choosing $n_\mu=2$ approximates each
quadrant (octant) with a single ray and also yields
$K_{ij}=1/3\delta_{ij}J$.  The algorithm is well-defined and unique
only for $n_\mu \le 12$ \citep{Bruls:1999}, corresponding to $n_{\rm
  a}=168$ (84 in two dimensions).  This should not be prohibitive for
the problems of interest.

\begin{figure}
\includegraphics[width=0.47\textwidth]{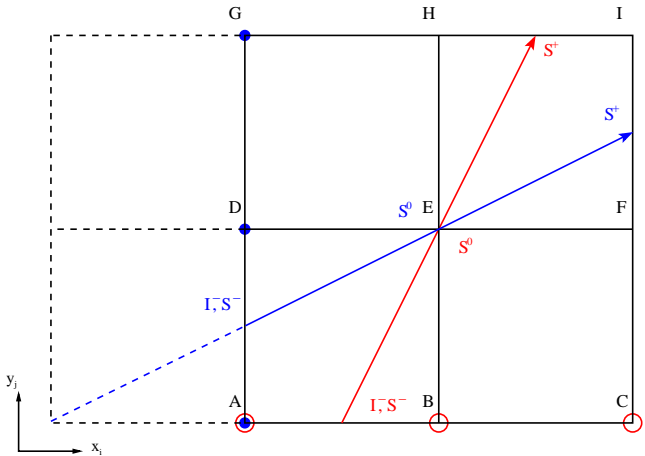}

\caption{
Schematic of the RT solution for an individual gridzone whose cell
center correspond to vertex E in a two-dimensional radiation grid.  In
this case $S^0$, and $\chi^0$ are known at E, and $I^0$ is to be computed.
$I^0$ is computed along a each ray using equation (\ref{eq:rtdisc}). Since
quantities $I^-$, $S^\pm$, and $\chi^\pm$ do not correspond to vertices
of the grid, they must be computed via interpolation from neighboring grid
zones.  The red ray intersects rows between vertices A and B (upwind)
or H and I (downwind). Hence the values of $S$, $\chi$ and $I$
at these vertices are used to interpolate $S^\pm$, $\chi^\pm$, and $I^-$.  Vertices C and G
are also used with quadratic interpolation.  A similar case holds
for the blue ray, but with interpolation performed on columns A-D-G
and C-F-I.  The open and closed circles denote vertices which are used
for the interpolation of $I^-$.  The dashed curve is an extension of the
blue ray which intersects A-B-C row. 
\label{f:zone}}
\end{figure}

For each ray $\hat{n}_k$, we compute a vector of direction cosines
$(\mu_{0k},\mu_{1k},\mu_{2k})$ with $\mu_{ik}=\hat{n}_k \cdot
\hat{x}_i$ and quadrature weights $w_k$. Then equations
(\ref{eq:J})-(\ref{eq:K}) become
\begin{eqnarray}
J &=& \sum_{k=0}^{n_a-1} w_k I_k \label{eq:Jsum}\\ H_i &=&
\sum_{k=0}^{n_a-1} w_k I_k \mu_{ik} \label{eq:Hsum}\\ K_{ij} &=&
\sum_{k=0}^{n_a-1} w_k I_k \mu_{ik} \mu_{jk}, \label{eq:Ksum}
\end{eqnarray}
where $I_k \equiv I(\hat{n}_k)$.

\subsection{Implementation of the Short-Characteristics Algorithm}
\label{shortchar}

The short characteristic method
\citep{Mihalas:1978,Olson:1987,Kunasz:1988} has been discussed
previously by several authors.  The basic computation step for a
single gridzone in both LTE and non-LTE problems is illustrated in
Figure \ref{f:zone} for the two-dimensional case.  Fluid radiative
properties and radiation variables (e.g. $\chit$, $B$, $\epsilon$,
$I_k$, $J$, $S$) are defined on a radiation grid.  The vertices of
this grid correspond to the cell centers of the MHD domain so that
fluid radiative properties are computed directly from the cell
centered MHD fluid variables.  Generalizing to three dimensional
domains is straight-forward.

At each vertex, the specific intensity $I^0_k$ at $\mathbf{x}^0$
is computed along each ray $\hat{n}_k$ from $\mathbf{x}^-_k$ to
$\mathbf{x}^+_k$.   For second-order interpolation the intensity
is given by
\begin{eqnarray}
I^0_k & = &  I^-_k e^{(-\Delta \tau^-_k)} + 
\Psi^-_k S^-_k  + \Psi^0_k S^0 + \Psi^+_k S^+_k,
\label{eq:rtdisc}
\end{eqnarray}
where $\Psi^-_k$, $\Psi^0_k$, and $\Psi^+_k$ denote interpolation
coefficients which depend on the opacities $\chi^-_k$, $\chi^0$, and
$\chi^+_k$ through the optical depth intervals $\Delta \tau^-_k$ and
$\Delta \tau^+_k$.  

The form of the interpolation coefficients $\Psi^-_k$, $\Psi^0_k$, and
$\Psi^+_k$ depends on the interpolation method used.  The standard
expressions for second order interpolation are listed in equations
(7a)-(9c) of \citet{Kunasz:1988}.  One drawback of these expression is
that they are subject to overshoot where gradients in $S^\pm_k$ and
$\chi^\pm_k$ are steep.  Fortunately, these cases can be handled with
B\'{e}zier-type interpolation as described in \citet{Auer:2003} and
\citet{Hayek:2010}.  With B\'{e}zier-type interpolation schemes, one
can utilize a control point $S^{\rm c}_k=S^0 - 0.5\Delta \tau^-_k
(\partial S/\partial \tau)^0_k$ to determine if overshoots are present
in the standard second-order expressions.  If $S^{\rm c}_k < {\rm
  min}(S^-_k,S^0)$ or $S^{\rm c}_k > {\rm max}(S^-_k,S^0)$ overshoots are
present and alternative expressions are utilized.  Suitable choices
follow from setting $S^{\rm c}_k=S^-_k$ or $S^{\rm c}_k=S^0$.
\citet{Hayek:2010} provide the corresponding expressions for
$\Psi^\pm_k$ and $\Psi^0_k$ in their Appendix A.  Similar methods are also
used to compute intervals $\Delta \tau^\pm_k$ (see e.g. equation A.3 of
\citealt{Hayek:2010}).

For one-dimensional problems, $\mathbf{x}^-_k$ and $\mathbf{x}^+_k$
correspond to neighboring grid vertices $x_{i-1}$ and $x_{i+1}$.
Hence, $I^-_k=I_{i-1,k}$, which was just computed in the neighboring
zone while $S^\pm_k$, and $\chi^\pm_k$ can be computed directly from
hydrodynamics variables at $x_{i \pm 1}$. In multidimensional
problems, $\mathbf{x}^-_k$ and $\mathbf{x}^+_k$ no longer correspond
to vertices of the radiation grid and variables $I^-_k$, $S^\pm_k$,
and $\chi^\pm_k$ must be interpolated.  We implement and test both
first-order (linear) and monotonic second order (quadratic)
interpolation schemes \citep{Auer:1994}.  Both methods prevent
overshoots and enforce positivity of the interpolants.  The choice is
particularly relevant for $I_k$, as second-order methods generally
produce much less diffusion of the radiation beam.  A drawback of
second order interpolation is that it places additional constraints on
the order in which one sweeps through gridzones and the stencil used
for the evaluation of $I_k$.

\begin{figure}

\includegraphics[width=0.47\textwidth]{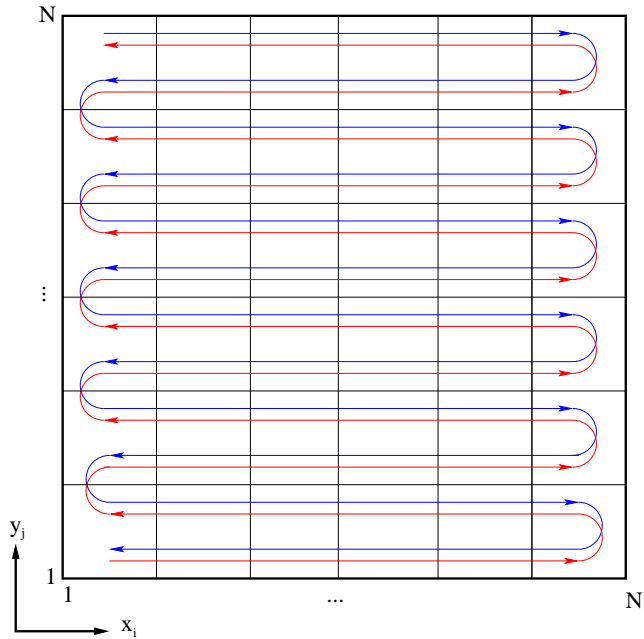}

\caption{
Progression of the sweep through a two-dimensional grid (domain or
subdomain) when linear interpolation is used.  The forward sweep (red
curve) first progresses parallel to $\hat{y}$, computing RT only for
upward pointing rays ($\hat{n} \cdot \hat{y} > 0$). For each row
(fixed $y_j$), one first sweeps parallel to $\hat{x}$, computing RT
along rays with $\hat{n} \cdot \hat{x} > 0$ until reaching the grid
edge $x_N$, then reverses direction and computes along rays with
$\hat{n} \cdot \hat{x} < 0$ until reaching the grid edge $x_0$.  This
continues until one reaches gridzone $(x_0,y_N)$.  The backward sweep
(blue curves) inverts the forward sweep, computing RT only for
downward pointing rays ($\hat{n} \cdot \hat{y} < 0$).  In the
Gauss-Seidel method, updated values of $S_{i,j}$ are incorporated into
the backward sweep, beginning with $S_{N,N}$. The three-dimensional
case is a straightforward generalization.
\label{f:sweep}}
\end{figure}

Consider the two rays depicted in Figure \ref{f:zone}.  We compute
interpolants $S^\pm_k$ and $\chi^\pm_k$ using known quantities at vertices
of the radiation grid.  If row A-B-C or column A-D-G correspond to
ghost (boundary) zones, $I_k^-$ can be computed from the (prescribed)
boundary intensities.  If they are not ghost zones, interpolation can
only be performed on zones in which $I_k$ has already been computed.  If
we first sweep along rows of fixed $y_j$ (as in Figure \ref{f:sweep}),
$I_k$ has only been computed at vertices A, B, C, and D.  This means
that $I_k$ is known for all vertices used in the linear interpolation of
$I^-_k$ as well as for quadratic interpolation (and any higher order
interpolation) of rays which intersect row A-B-C.

We refer to rays that intersect the column A-D-G, such as the blue one
in Figure \ref{f:zone}, as ``shallow'' rays.  Shallow rays are a
potential problem for quadratic (and higher order) interpolation,
since $I_k$ at G has not been computed. When quadratic or higher order
interpolation is desired, such rays can be handled in a number of
ways.  One possibility is to switch the order of the sweep for shallow
rays so that it first proceeds in the y direction along columns of
fixed $x_i$.  In this case $I_k$ for shallow rays will be known at
vertices A, B, D, and G.  The main drawback (discussed further in
Section \ref{nonlte} below) is that one is unable to implement a
Gauss-Seidel iteration for non-LTE problems.

One can also construct alternatives by extending the stencil beyond
vertices A-I.  For example, one can extend shallow rays until they
intersect row A-B-C as shown by the dashed curve in Figure
\ref{f:zone}.  A drawback of this solution is that it requires modest
additional effort for computing $\Psi^-_k$, although this can be
alleviated by computing only on the first iteration and reusing it for
subsequent iterations \citep[e.g.][]{Hayek:2010}.  Alternatively
quadratic interpolation could be preformed using A, D, and the vertex
directly below A \citep{Kunasz:1988}.

These two solutions share common drawbacks.  For parallelization with
domain decomposition, only one ghost zone is needed per grid zone on a
subdomain face, when only vertices A-I are used.  Extension of shallow
rays beyond this stencil requires the passing of additional data and
associated bookkeeping.  More philosophically, we feel it is desirable
to treat all rays as consistently as possible.  In either of these
schemes, RT along some rays will be computed using only neighboring
grid zones, while other rays will not. Our preference is to treat
all rays on the same footing.

For this reason, we have decided to switch the order of the sweep for
shallow rays.  Athena is implemented so that each sub-grid of the
domains has regular spacing and therefore gridzones with fixed aspect
ratio.  This means that the distinction between rays that are shallow
and those that are not is equivalent for each grid zone.  However, our
definition of a shallow ray depends upon the direction of the sweep.
The blue ray in Figure \ref{f:zone} is shallow because we first
traverse the grid along rows of fixed $y_j$, only moving to $y_{j+1}$
when intensity has been computed for all gridzones in the row $y_j$,
as depicted in Figure \ref{f:sweep}.

If we reverse the sweep so that we first traverse columns of fixed
$x_i$, the blue ray will no longer be shallow, as the intensity at G
will be computed before it is needed for the computation of the
intensity at E.  In this case the red ray is now a shallow ray as the
intensity at C will not have been computed before it is needed to
compute the intensity at E.  Hence, by varying the sweep direction, we
can handle all rays and accommodate a quadratic interpolation scheme
which computes all intensities in a gridzone $(x_i,y_j)$ only using
intensities from neighboring gridzones $(x_{i \pm 1},y_{j \pm 1})$.

\subsection{Iterative Methods for Non-LTE Problems}
\label{nonlte}

We now describe how we handle non-LTE problems iteratively.  Following
common convention we denote the angle averaged formal solution of
the RT equation (hereafter, simply the formal solution) in operator
notation as
\begin{equation}
J = \Lambda S.
\label{eq:formalsol}
\end{equation}
Here, $\Lambda$ is a linear operator representing the (discretized) formal solution,
and $J$ and $S$ are vectors spanning each gridzone in the simulation
domain.  
Using equation (\ref{eq:sourcefunc}) to eliminate $J$, one obtains
an equation for $S$ in terms of $B$
\begin{equation}
S = (1 -\epsilon)\Lambda[S]+\epsilon B.
\label{eq:sourcefunc2}
\end{equation}
Since  $\Lambda$ is a linear operator we can solve for $S$
\begin{equation}
S = [1 - (1 -\epsilon)\Lambda]^{-1}[\epsilon B].
\label{eq:sourcefunc3}
\end{equation}

If one can invert $\Lambda$ a formal solution of the non-LTE
problem follows from solving (\ref{eq:sourcefunc3})
and obtaining $J$ from (\ref{eq:formalsol}).  However, for three-dimensional
problems $\Lambda$ is a very large matrix and not sparsely populated when
systems are far from LTE so its direct inversion is impractical. Therefore,
equation (\ref{eq:sourcefunc2}) is usually solved via iteration.

A simple iterative scheme for solving equation (\ref{eq:sourcefunc2})
begins with an initial guess for the source function $S^N$, which is then
used to compute an improved estimate $S^{n+1} = (1
-\epsilon)\Lambda[S^n]+\epsilon B$.  However, this method (often
referred to as Lambda Iteration) has very poor convergence properties.
For practical problems, ALI
methods \citep{Cannon:1973} are commonly
used. \citet{Rybicki:1991}, \citet{Hubeny:2003} and TF95 provide
useful reviews of ALI methods and we refer the reader to these works
for a more in-depth discussion.  Here we just summarize the basic
concepts involved.

In ALI methods one solves equation (\ref{eq:sourcefunc3}) directly,
but using an approximate form $\Lambda^*$ which is easier to invert
then the full $\Lambda$ operator.  Since only the approximate
$\Lambda^*$ is used, iteration is still necessary.  Numerous choices
for $\Lambda^*$ have been proposed, but it has been argued that simply
taking the diagonal elements of the full $\Lambda$ matrix represents a
near-optimal choice \citep{Olson:1986}. \citet{Olson:1987} provide
expressions for diagonal elements of $\Lambda$ when short
characteristics are used.  In each grid zone the change in the source
functions $\Delta S_i= S^{n+1}_i - S^n_i$ can be written as
\begin{equation}
\Delta S_i = \frac{(1-\epsilon_i) J^n_i+\epsilon_i B_i - S^n_i}
{1-(1-\epsilon_i)\Lambda_{ii}},
\label{eq:dsjacobi}
\end{equation}
where the subscript $i$ enumerates all gridzones in the domain.

As TF95 discuss, when $J^n$ is exclusively used in equation
(\ref{eq:dsjacobi}), the ALI scheme is equivalent to the Jacobi
iterative method for solving linear systems. TF95 show that one can
construct a Gauss-Seidel algorithm by incorporating the new values of
$J^{n+1}_{i'<i}$ in equation (\ref{eq:dsjacobi}) as these become
available. Here $i' < i$ refers to gridzones where $J$ has already
been updated.  The complexity of devising a Gauss-Seidel algorithm for
RT comes from the fact that the computation of specific intensity
$I_{i,k}$ for some subset of the rays $\hat{n}_k$ need to be computed
using $S^{n}_{i'<i}$ rather than $S^{n+1}_{i'<i}$ (i.e. old rather
than new values of the source function).  Therefore the contribution
from these particular rays to $J^{n+1}_{i'<i}$ must be corrected as the
updated values $S^{n+1}_{i'<i}$ become available.

TF95 give a detailed description of how to implement such an algorithm
on a one-dimensional domain.  The algorithm requires storing a modest
amount of data in each gridzone, but very little additional
computation.  The convergence rate is improved by a factor of two, so
problems requiring several iterations gain nearly a factor of two
decrease in computational effort for only a minor increase in code
complexity.

When linear interpolation is used, the generalization of their
one-dimensional method to two and three-dimensional domains is
straight-forward.  The two-dimensional sweep proceeds as depicted in
Figure \ref{f:sweep}.  The vertices in the radiation grid correspond
to cell centers $(x_i,y_j)$.  The sweep generally proceeds with $i$ as
the more rapidly varying index.  Consider a domain with $N_x=N_y=N$
for simplicity.  In each gridzone $(x_i,y_j)$, we first compute the
intensity $I_k$ for all upward directed rays $(\hat{n}_k \cdot \hat{y}
> 0)$ in the forward sweep and then for all downward directed rays
$(\hat{n}_k \cdot \hat{y} < 0)$ on the reverse sweep.

On the reverse sweep, the upper right gridzone $(x_{N},y_{N})$ is the
first in which the computation of all new intensities $I^{n+1}_k$ is
completed.  At this point $J^{n+1}_{N,N}$ is completely specified and
we compute $S^{n+1}_{N,N}$.  From here on, all subsequent RT
computations use $S^{n+1}_{N,N}$ rather than $S^{n}_{N,N}$.  However,
this alone is not sufficient to make it a Gauss-Seidel scheme, because
the contributions to $J^{n+1}_{N-1,N}$, $J^{n+1}_{N,N-1}$, and
$J^{n+1}_{N-1,N-1}$ from upward directed rays on the forward sweep
used $S^{n}_{N,N}$.  These must also be updated using $\Delta S_{N,N}
= S^{n+1}_{N,N}- S^{n}_{N,N}$ and weights which were saved on the
forward sweep.  We also update the outgoing intensities $I^{n+1}_k$
(since they were also computed using $S^{n}_{N,N}$) as they correspond
to the incoming intensities in neighboring gridzones.  Since the
corresponding weights have already been computed as part of the
forward sweep, the additional computational cost is very modest.

Following the discussion in Section \ref{shortchar}, we note that
feasibility of performing a Gauss-Seidel iteration with quadratic
interpolation is dependent on the way shallow rays are handled.
Reorienting the sweep for shallow rays so that $j$ is more rapidly
varying index, but keeping $i$ as the rapidly varying index for
remaining rays does not allow for an efficient Gauss-Seidel scheme
because some of the necessary $J^{n+1}$ (and therefore $S^{n+1}$) are
not available when the backward sweep begins.\footnote{For
  two-dimensional domains one can devise an efficient Gauss-Seidel
  algorithm that sweeps diagonally through the grid, but this
  implementation does not generalize to three dimensions.}  In the
light of this issue, we have implemented Gauss-Seidel routines only
with linear interpolation.  For problems where quadratic interpolation
is preferable, we default to the Jacobi method (i.e. standard ALI).

We continue the iteration until some convergence criterion is met.
Consistent with previous work, we stop iterating when the maximum
relative change in the source function over the whole domain is less
than some prescribed threshold $\delta_c$
\begin{equation}
\max \left(\frac{|\Delta S_i|}{S_i}\right) \le \delta_c.
\label{eq:conv}
\end{equation}
For LTE problems that use iteration to handle boundary conditions, $S$
does not change from one iteration to the next and we replace $S_i$
with $J_i$ in equation (\ref{eq:conv}).

The choice of $\delta_c$ is clearly an important input to the method,
but there is no firmly established criterion and the optimal choice
depends on a number of considerations that may be problem dependent.
Since the computational cost of the method generally scales linearly
with the number of iterations performed and a lower threshold leads to
more iterations, there is a tradeoff between accuracy and
computational expediency.  With the exception of the uniform
temperature non-LTE atmosphere, the tests presented in section
\ref{tests} were performed using $\delta_c=10^{-5}$.  Increasing
$\delta_c$ to $10^{-4}$ had a negligible impact on the linear wave
tests.

Our expectations based on the tests we have performed so far are that
for the problems of primary interest to us (e.g. shearing box
simulations of accretion disks) $\delta_c \sim 10^{-5} - 10^{-3}$ will
be sufficient, consistent with studies using similar methods
\citep[e.g.][]{Hayek:2010}.  However, we emphasize that the
appropriate choice will be problem dependent and must be assessed on a
case-by-case basis. We view the choice of $\delta_c$ in roughly the
same terms as we view the choice of grid resolution.  One can adopt a
threshold based on previous results and experience, but ultimately
one needs to compute the problem using a range of $\delta_c$ and choose
a sufficiently small value such that the results are insensitive to
the choice.

\subsection{Boundary Conditions and Parallelization}
\label{boundcond}

Boundary conditions and domain decomposition in Athena are both
implemented for MHD via the use of ghost zones, and we implement RT boundary
conditions in an analogous way.  The solver computes RT in gridzones
on a boundary (domain or subdomain) in the same way as an interior
gridzone, but using the intensities and source functions from the
ghost zones to compute the relevant integration weights and
interpolants.  The intensities and source functions in the ghost zones
are determined according to prescribed boundary conditions.

In general, boundary conditions for the MHD integrator will not
translate directly to boundary conditions for the RT solver.
Different problems with the same MHD boundary conditions may require
different boundaries for the radiation field. Hence separate boundary
conditions must be prescribed when using the RT solver.  For the code
test problems presented in section \ref{tests}, we have implemented
two types of boundary conditions specifying either fixed incident
intensity or periodic intensities on the boundaries.  Other boundary
conditions can be specified via user defined functions.

Athena runs on parallel machines using domain decomposition
implemented through MPI calls.  The MHD integrator passes all
conserved variables and passive scalars from faces of neighboring
subdomains to ghost zones.  The MHD integrator requires either four or
five ghost zones for each gridzone on the subdomain face.  The RT
solver operates analogously, passing intensities and source functions,
but only requires one ghost zone for each gridzone on a subdomain face.

The main differences between the RT solver and the MHD integrator are
the frequency and quantity of data that must be passed.  For each
frequency bin in every ghost zone $I$ must be passed for all $n_a$ rays
with quadratic interpolation or, alternatively, $n_a/2$ incoming rays
with linear interpolation.  For non-LTE problems $S$ and $\mathbf{H}$
must also be passed.  Hence for quadratic interpolation, the code
passes a total of $n_f (n_a+1+n_{\rm dim})$ floating point variables
per face gridzone per {\it iteration}, where $n_{\rm dim}$ is the
number of dimensions in the domain.  In contrast, the MHD integrator
typically passes $\sim 50$ floating point variables per face gridzone
per {\it timestep}.  For problems where $n_a$ and $n_f$ are small and
few iterations are required (e.g. an LTE grey problem), the volume of
RT data is therefore comparable to and may even be less than the
amount of data passed by the MHD integrator.

We note that the use of iteration to handle subdomain boundary
conditions may lead to some dependence on the number of subdomains
that are used.  We have considered the sensitivity of our results to
this issue by performing most of the tests described in section
\ref{tests} both with and without domain decomposition.  In practice,
the converged mean intensities do not differ (relative to the non
decomposed domain) by more than $\sim \delta_c J$.  The sensitivity is
highest for problems where the optical depth across an individual
subdomain is of order unity or smaller, Problems with optically thick
subdomains generally lead to smaller discrepancies.  Since we already
choose our convergence criterion to be at a level that minimizes the
impact on our results, this sensitivity to the domain decomposition
should not lead to significant errors.

\section{Interface of the Radiative Transfer Solver to the MHD Integrator}
\label{interface}

There are two regimes in which the effect of radiation on the MHD
is important.  The first is when the radiation field is a significant
contribution to both the energy and momentum fluxes in the flow.  In this
regime, the radiation source terms in the MHD equations can be very stiff,
and the equations contain wave modes which propagate at close to the speed
of light.  Both of these properties require significant modification to
the underlying MHD integrators in order to enable accurate and stable
integration.  In JSD12 we describe a method for this regime based on an
extension of the modified Godunov method of SS10 to multidimensions, with
a VET (defined as ${\sf f} = {\sf P_{\rm rad}}/E_{\rm rad}$) computed
from a formal solution of the RT equation using the module described
in this paper.  At each time step, the RT solver computes the radiation
field as described in Section \ref{radtrans}, evaluating ${\sf K}$ and
$J$ via equations (\ref{eq:Jsum}) and (\ref{eq:Ksum}).  We the compute
the  VET using ${{\sf f}=\sf K}/J$ as described in section 3.4 of JSD12.

The second regime is when the radiation pressure can be ignored, and
the effect of radiation is only through the heating and cooling source
terms in the energy equations.  In principle, the modified Godunov
method adopted in JSD12 would be an attractive approach for handling
the stiff energy source term that can arise in this regime as well.
However, the modified Godunov method requires that one compute the
gradient of radiation source terms on the plane of primitive variables.
This in turn requires analytic expression for the radiation sources in
terms of the fluid variables.  Hence, it is generally not a viable method
for problems where the radiation properties are complicated functions
of frequency and fluid variables, as may be the case with bound-free
and bound-bound atomic opacities or Compton scattering.

These limitations motivate us to implement an alternative method to
directly compute the radiation source term in the fluid energy
equation (\ref{eq:energy}) and couple it to the standard MHD
integrators.  When operating in this mode, we perform the formal
solution at the beginning of each timestep.  We first compute fluid
radiation properties in each gridzone $\mathbf{x}_i$ of the domain.
This includes the variables $\chit_i$, $B_i$, and $\epsilon_i$, which
are computed via user-defined functions of the conserved MHD variables
and passive scalars from the previous timesteps.  We use these, along
with $J_i$ from the previous time-step, to initialize $S_i$.
Once the formal solution is completed, we account for the source
function on the right hand side of equation (\ref{eq:energy}) via an 
operator split update of $E$. We first compute the radiative source
function in each zone and then update the total energy 
\begin{equation}
\Delta E_i =  \delta t(Q_{\rm rad})_i.
\end{equation}
The standard MHD integration algorithm then proceeds using this
``new'' value for $E_i$.

We compute $Q_{\rm rad}$ in one of two ways, depending on
the characteristic optical depth.   We either use the integral
form
\begin{equation}
Q^{\rm int}_i = 4 \pi \chit_i (J_i - S_i)=4 \pi \chia_i (J_i - B_i),
\end{equation}
or the differential form
\begin{equation}
Q^{\rm dif}_i = -4 \pi \mathbf{\nabla} \cdot \mathbf{H}_i,
\end{equation}
Previous work \citep[ and references therein]{Bruls:1999} has
demonstrated that the integral form is inaccurate when the optical
depth per gridzone is large. In this case $J_i-B_i \ll B_i$ while
$\chia_i$ is large so round-off errors can be greatly amplified.  The
integral form, however, is more accurate when $\chia_i \Delta
\mathbf{x}_i \lesssim 1$ \citep{Bruls:1999}.

Therefore, we have designed our RT solver to compute either form
of $Q_{\rm rad}$, depending on the regime of the computation.  In most
applications of interest, there is a transition from optically thick
to optically thin regions, so we must specify a criterion for switching
between the differential and integral forms in the same domain.  For the test
problems considered here, we find a simple switch 
\begin{displaymath}
(Q_{\rm rad})_i = \left\{ \begin{array}{ll}
Q^{\rm int}_i & {\rm if} \;  \chit_i \Delta x_i \le 1\\
Q^{\rm dif}_i & {\rm otherwise}
\end{array} \right.
\end{displaymath}
to be sufficient.  This has the advantage that it is a purely local
criterion.  Using a method which more smoothly interpolates between
the two regimes \citep[see e.g.][]{Hayek:2010} did not improve performance
in a measurable way, but may be preferable for more sophisticated
applications.

Due to the explicit update, we must take care in choosing a time
step. In the absence of radiation the MHD integrator chooses a time
step $\delta t_{\rm C}$ based on the CFL constraint.  In principle,
this time step can be much larger than the radiative cooling time,
which could lead to obvious errors, such as the energy density
becoming negative.  As we elaborate upon in section \ref{linwave}, one
can derive a generalized CFL condition for a radiating fluid based
on the need to resolve the damping time for a non-equilibrium
radiation diffusion mode.  This time scale $\delta t_{\rm rd}$ is
generally most restrictive when the optical depth per gridzone $\chit
\Delta x \sim 1$, in which case
\begin{equation}
\delta t_{\rm rd} \approx  \frac{E_{\rm gas}}{E_{\rm rad}}\frac{a}{c} \delta t_{\rm C},
\end{equation}
assuming the adiabatic sound speed $a$ (rather than the Alfv\'{e}n
speed) sets the CFL condition.
This generalized CFL
  constraint can be quite stringent, requiring short time steps and
  increasing computational costs if either $a \ll c$ or $E_{\rm rad}
  \gtrsim E_{\rm gas}$.  Hence, many problems will require the use of
  the VET method described in JSD12, which uses timesteps determined
  by the standard (non-radiative) CFL constraint. In practice, we are almost
  always limited to problems with $E_{\rm gas} < E_{\rm rad}$, so we do not
  attempt to include the radiation momentum source term in equation
  (\ref{eq:mom}) as it is generally small for problems that are computationally
  feasible with operator splitting.

The algorithm described above will, in general, only be first order
convergent.  Note that we could construct a second-order scheme when
using Athena's VL+CT integrator \citep{Stone:2009}, by performing the
operator split update before the corrector step in the
predictor-corrector scheme.  However, some of the advantages of the
second order convergence will be lost due to the increased diffusivity
of the VL+CT relative to the CTU+CT scheme \citep{Gardiner:2008}.
Hence, we have not yet pursued the possibility although it may prove
to be a useful avenue for future work.

\section{Tests}
\label{tests}

Our test problems fall into roughly two categories: stand-alone tests
of the RT solver on fixed domains and tests of the coupled MHD
integrator and RT solver in fully time dependent calculations.  The
former are particularly useful for evaluating the RT solvers
performance on multidimensional and non-LTE problems.  For the latter,
we focus primarily on simpler LTE problems, so we can compare the
simulations result directly to precise analytic or semi-analytic
solutions.

Further tests of the RT solver as part of the VET method are presented
in JSD12.

\begin{figure}
\includegraphics[width=0.47\textwidth]{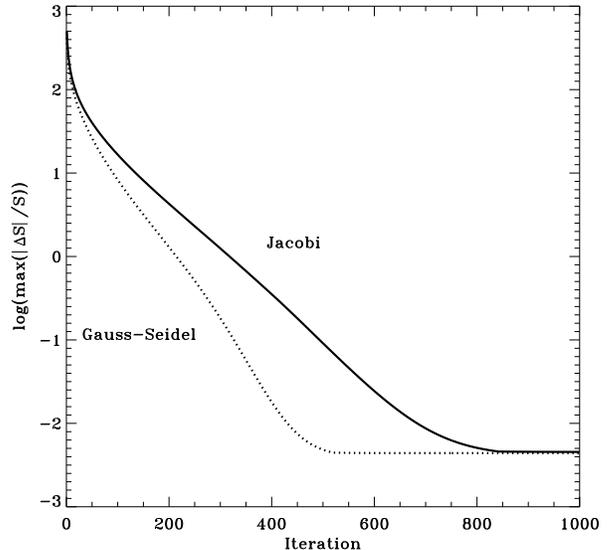}

\caption{
Convergence of the ALI methods on a highly non-LTE uniform
atmosphere with $\epsilon=10^{-6}$.  The curves show the maximum
relative difference between the numerically computed $S$ and the
analytic solutions implied by equation (\ref{eq:jana}) for the Jacobi
(solid) and Gauss-Seidel (dotted) methods. We compute the numerical
solutions using a one-dimensional domain with 9 gridzones per decade
in optical depth.
\label{f:conv_comp}}
\end{figure}

\subsection{Uniform Temperature Non-LTE Atmosphere}

We begin by solving the monochromatic RT problem in a uniform
temperature, one-dimensional scattering dominated atmosphere.  This
test is particularly useful for evaluating the RT solver's performance
on a non-LTE systems and evaluating the convergence properties of
Jacobi and Gauss-Seidel iterative schemes.  We adopt the two-stream
approximation for the RT solver so we can compare directly with analytic
solutions based on the Eddington approximation.  Since we assume a
uniform opacity $\kappa$ and temperature $T$, the analytic solution is
only a function of optical depth $d \tau = \chi dz$, the thermal
source function $B$ and photon destruction probability $\epsilon$.
With these assumptions the mean intensity $J$ is given by
\begin{equation}
J=B\left(1-\frac{e^{-\sqrt{3\epsilon}\tau}}{1+\sqrt{\epsilon}}\right).
\label{eq:jana}
\end{equation}
We assume that $\chi \propto \rho$ and $\rho$ increases exponentially
(but keep $\epsilon$ constant) with distance from the upper boundary,
which has no incoming intensity.  This provides an exponential
variation in $\tau$ which is well-suited for resolving the transition
from LTE to non-LTE within the atmosphere.

Figure \ref{f:conv_comp} shows the convergence of the true error of
the numerically derived solutions.  This is evaluated as the maximum
relative difference $|\Delta S|/S$, with $\Delta S$ the difference of
the numerically derived $S$ from the analytic solution.  We first
consider a one-dimensional domain with $\epsilon=10^{-6}$, as this
gives a highly non-LTE atmosphere and facilitates direct comparison
with Figure 3 in TF95.  We initialize the radiation field to be in LTE
everywhere ($J=B$).  We consider two different iterative schemes:
Jacobi and Gauss-Seidel.  As expected, the convergence rate of the
Gauss-Seidel methods is nearly a factor of two better than Jacobi.  We
assume nine gridzones per decade in $\tau$ to match TF95 and our
convergence rates agree reasonably well with those shown their Figure
3.  

\begin{figure}
\includegraphics[width=0.47\textwidth]{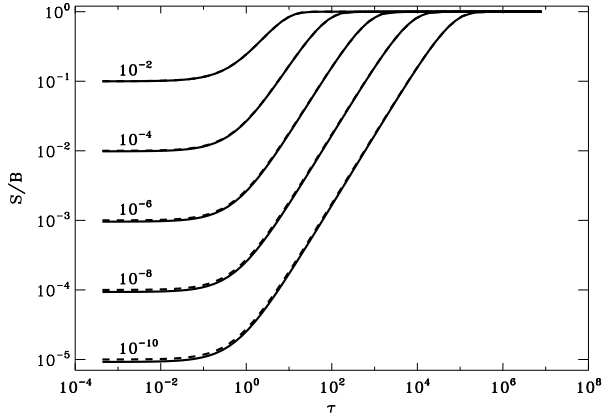}

\caption{
Comparison of numerical (solid) and analytic (dashed)
solutions of the source function monochromatic, uniform non-LTE
atmospheres as a function of optical depth.  Each set of curves
corresponds to a different photon destruction parameter, running from
$\epsilon=10^{-2}$ to $10^{-10}$ from top to bottom.  We compute
the numerical solutions simulations using cubic three-dimensional domains
with $64^3$ gridzones, distributed over 64 MPI subdomains.  The optical
depth variation is aligned with the $z$-axis of the simulation domain and
the solution is uniform in the horizontal directions.
\label{f:eps_comp}}
\end{figure}

We have also implemented the successive over-relaxation (SOR) method
of TF95, and find rapid convergence, consistent with that shown in
Figure 3 of TF95.  We have tested SOR on both one-dimensional and
two-dimensional domains and find that it is an effective method as
long as all boundary intensities are fixed during iteration.  However,
if the intensities on one of the boundaries vary from one iteration
step to the next, the method is generally not stable. For example,
instability occurred when we used periodic boundary conditions or when
we employed subdomain decomposition.  Since most of our primary
science goals involve problems that require the use of periodic
boundary conditions or domain decomposition, we do not consider SOR a
generally viable method for our work.  Nevertheless, it may be an
effective method for a modest sized problem that can be run serially
with fixed boundary intensities.

We next consider the same test problem, but use a cubic
three-dimensional domain with $n_\mu=2$.  We align the variation of
density with the $z$ axis of the domain and use periodic boundaries in
the horizontal directions.  Figure \ref{f:eps_comp} shows a comparison
of the numerical and analytical solutions for various choices of
$\epsilon$.  The agreement between the numeric and analytic solutions
is quite good overall, but tends to be poorest at low optical depths.
For fixed resolution, the discrepancies with the analytic solution
tends to increase as $\epsilon$ decreases and the domain deviates more
strongly from LTE.  The accuracy of the numerical solution improves
with increasing resolution, but the number of iterations needed for
convergence increases roughly linearly with resolution.  The number of
iterations required for convergence also increases as $\epsilon$
decreases.  Hence, greater deviations from LTE require a greater
number of iterations for convergence, as one would expect.

\begin{figure}
\includegraphics[width=0.47\textwidth]{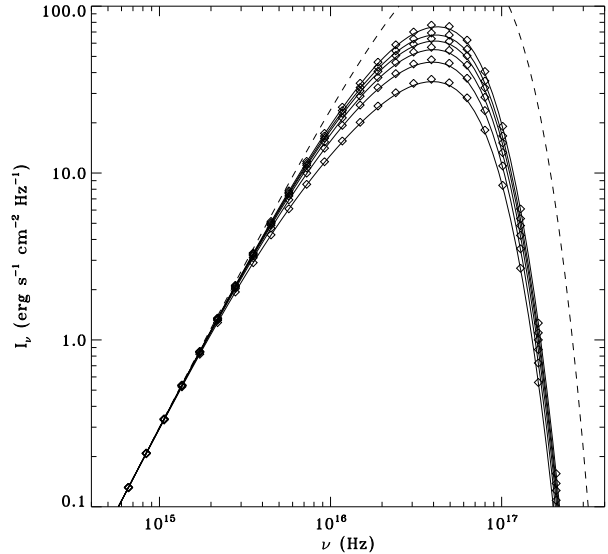}

\caption{
Comparison of Athena (diamonds) and Feautrier (solid) emission spectra
from the upper boundary of a one dimensional atmosphere. Both
calculations assume isotropic electron scattering and free-free
(Bremsstrahlung) absorption and emission for a completely ionized H
plasma.  The intensities are computed using the same angular grid
corresponding to abscissas of a 16 point Gauss-Legendre quadrature of
the interval (1,1). The plotted intensities (from top to bottom)
correspond to $\cos i =$0.10, 0.28, 0.46, 0.62, 0.76, and 0.99.  The
atmospheres have constant temperature ($10^6$ K) and density which
varies exponentially with distance, rising from $10^{-6}$ g/$\rm cm^2$ at
the upper (surface) boundary to $10^{-4}$ g/$\rm cm^2$ and lower boundary.
For comparison, we plot the corresponding blackbody at $10^6$ K as a
dashed curve.
\label{f:spec}}
\end{figure}

Although some RT problems do require explicit frequency coupling
(e.g. Compton scattering, partial redistribution), many problems can
be treated in the approximation that frequencies are not explicitly
coupled.  Multifrequency problems are then just a series of single
frequency calculations and, hence, a straightforward generalization of
the monochromatic problem.  Figure \ref{f:spec} shows the intensity
spectrum from a multifrequency calculation done with Athena for a
uniform temperature atmosphere.  We again assume $\rho$ varies
exponentially with distance, rising from $10^{-6}$ g/$\rm cm^2$ at the
upper boundary to $10^{-4}$ g/$\rm cm^2$ and lower boundary.  The
results plotted here are for $N_x = 256$.

\begin{figure*}
\begin{center}
\includegraphics[width=0.7\textwidth]{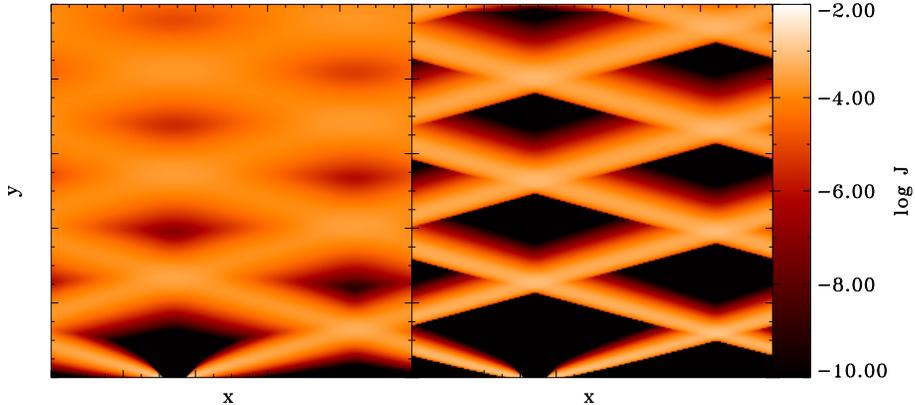}
\end{center}
\caption{
Mean intensity $J$ due to incident beams propagating through a
rarefied, two-dimensional domain.  The source function and opacity are
zero everywhere and the horizontal boundaries are periodic.  The
boundary condition at the top and bottom of the domain are zero
incident intensity except for two gridzones at the base of the domain.
In each of these gridzones $I$ is non-zero for a single ray.  The two
rays make the same angle with the y-axis, but are oppositely directed
in x, with $\hat{n} \cdot \hat{y} = 0.174$ and $\hat{n} \cdot \hat{x}
= \pm 0.628$.  The left and right panels show results from
computations with linear interpolation and monotonic quadratic
interpolation,respectively.
\label{f:beam2d}}
\end{figure*}

Incoming intensity at the upper boundary is assumed to be zero and
$I_\nu=B_\nu$ at the lower boundary.  We include isotropic electron
scattering opacity and free-free (Bremsstrahlung) emission and
absorption.  The electron scattering is modeled as isotropic and the
cross-section is the Thomson cross-section.  For simplicity free-free
processes are computed assuming a Gaunt factor of unity.  The plasma
is assumed to be completely ionized Hydrogen.  Hence, $\epsilon_\nu$,
$B_\nu$, and $\chi_\nu$ are all functions of frequency.  However, for
an individual frequency the calculations are very similar to those
described above.  The only difference is that $\epsilon_\nu$ is now a
function of depth as well, due to the different dependence of
scattering and absorption opacity on $\rho$.

Figure \ref{f:spec} also shows the results of a Feautrier calculation
for the same atmosphere using the same angular grid.  The two
calculations generally agree quite well, although there is a tendency
for the Athena solver to give slightly higher intensities for
frequencies where the spectrum deviates from blackbody.  The
discrepancy between the results is a function of spatial resolution
with agreement between the two codes improves as the $N_x$ is
increased in the Athena calculation.  Calculations on two and three
dimensional domains (but with density varying only in one dimension)
yield similar results.

\subsection{Beam Tests in Two Dimensions}

We now consider the propagation of crossing beams of radiation,
incident on the boundary of a rarefied ($B=0$, $\chi=0$), periodic
domain.  This test is particularly useful for evaluating the amount of
diffusion associated with the interpolation schemes for the specific
intensity.  It is also useful testing the performance of periodic and
subdomain boundary conditions.

The results for a two-dimensional domain with periodic boundary
condition in the horizontal direction are shown in Figure
\ref{f:beam2d}.  The figure compares a computation with linear
interpolation to one with quadratic monotonic interpolation.  Our
implementation of these methods is described in Section
\ref{shortchar}.  It is clear from Figure \ref{f:beam2d} that linear
interpolation leads to substantially greater diffusion of the
radiation beam.

Depending upon the application, the additional diffusion in the linear
interpolation scheme can be either advantageous or problematic.  On
one hand, a less diffusive scheme allows one to model important
effects, such as shadowing by optically thick material, with greater
fidelity.  Indeed, the ability to more accurately capture such effects
is an important motivation for using RT instead of more ad hoc closure
prescriptions, such as FLD.  

However, computational expedience limits the angular resolution we
can achieve.  When only a modest number of rays are used with a less
diffusive scheme, fan-shaped ``spokes'' can appear in the mean
intensities and Eddington factors, if the emission in a small number
of grid zones significantly exceed that of surrounding zones.  Indeed,
our short characteristics based method is not well suited to problems
with bright point sources for this reason, but even in applications
with distributed emission regions, there can be relatively confined
regions with larger than averaged emission (e.g. due to magnetic
dissipation).  In such cases, a greater degree of diffusion in the
intensity can mitigate unphysical effects which would otherwise arise
due to the limited angular resolution.

A related test of an RT routine is its ability to cast a shadow when
an optically thick obstruction is present in the domain.  We present
such a calculation in Section 5.5 of JSD12, where the ablation of an
optically thick cloud is studied.  In this case the RT solver was used
to compute the radiation field using linear interpolation for the
intensity field of neighboring zones.  Figure 14 of JSD12 demonstrates
that our RT solver can produce sharply defined umbra and penumbra
under such conditions.  FLD and other approximate moment methods
generally fail this test \citep[see e.g.][]{Hayes:2003}.

\begin{figure*}

\begin{center}
\includegraphics[width=0.7\textwidth]{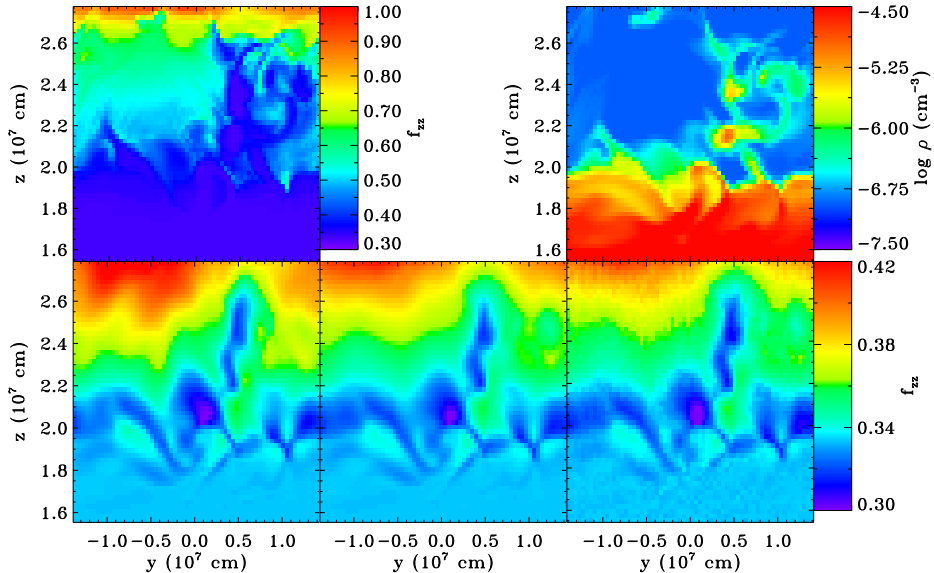}
\end{center}

\caption{
Comparison of Eddington factors computed using our RT scheme with MC
and FLD computations in a representative two-dimensional slice near
the top of a three-dimensional domain.  We plot the $f_{zz}$ component
of the Eddington tensor for computations with our Athena solver using
either 24 (bottom, left) or 168 (bottom, center) angle bins, and for the MC
(bottom, right) and FLD methods (top, left).  We also plot $\rho$ in
the same slice for comparison (top, right).  Note the larger range
for the color bar in the panel showing the Eddington factor for the
FLD computation.
\label{f:fldmc}}
\end{figure*}

\subsection{Comparison with Monte Carlo and FLD Methods}
\label{fldmc}

We now focus on comparing the performance of our short characteristics
solver (referenced throughout this section as the SC method) with two
alternative methods: FLD and Monte Carlo (MC).  Our motivation is
two-fold: in part, we want to evaluate the performance on a fully
three dimensional domain, but impose as few restrictive assumptions
(e.g. the Eddington approximation) on the radiation field as
possible. Since there is a paucity of such truly three dimensional
problems with analytic solutions, comparison with alternative RT
solution methods is the best alternative.  In addition, FLD and MC
methods are, in principle, some of the most computationally efficient
alternatives to short characteristics solvers, so direct comparison
may allow us to assess the relative merits of different methods.

For this comparison we use a three dimensional snapshot from a
stratified shearing box simulation, corresponding to a gas pressure
dominated patch of an accretion disk \citep{Hirose:2006}.  This
simulation was computed with the Zeus MHD code, using the FLD solver
developed by \citet{Turner:2001} and subsequently modified by
\citet{Hirose:2006}.  They solved the radiation moment equations using
a flux limiter of the type described in \cite{Levermore:1981}. Further
details about the particular snapshot used here can be found in
\citet{Blaes:2006}.  From the $E_{\rm rad}$ dump, we compute $F_{\rm
  rad}$ and the Eddington factor $\sf f$, using finite differences and
flux limiters consistent with those employed in the numerical
simulation.

We solve the RT equation on this snapshot using both our SC solver and
the MC code described in \citet{Davis:2009}.  For both calculations, we
assume isotropic electron scattering and monochromatic RT (i.e. a
single frequency bin) with mean opacities equal to those used in the
Zeus simulation ($\chia =3.7 \times 10^{53} \rho^{11/2} E_{\rm
  gas}^{7/2}$ and $\chis=0.33 \rho$, both in cgs units).  We assume no
incoming intensity at the surface boundaries and periodicity in the
horizontal directions.  The latter assumption is inconsistent with the
use of shearing periodic boundaries in the radial direction in the
numerical simulation, but this does not contribute significantly to
the discrepancy between the SC and FLD methods\footnote{We have
  implemented shearing boundaries in our SC solver and confirmed
  this. We show the results from the SC solver with periodic boundary
  conditions to facilitate comparison with the MC calculation which do
  not support shearing periodic boundaries.}

We compare the radiation moments (${\sf P}_{\rm rad}$, $\sf f$,
$\mathbf{H}_{\rm rad}$, and $E_{\rm rad}$) output by the SC/MC
solvers with those determined by the FLD method.  Independent of the
variable used for comparison, we find reasonable agreement between the
SC and MC solvers, but discrepancies with the FLD approximation.  For
brevity we will focus on a single scalar quantity, $f_{zz}$, since it
characterizes the variation of angular distribution of the
radiation field across methods.

Figure \ref{f:fldmc} shows a comparison of $f_{zz}$ among the
various methods for a representative two-dimensional slice near the
top boundary of the simulation domain.  In the top row, the left and
middle panels show results from the SC solver, using 24 and 168
angles, respectively.  The top right panel shows the MC results and
the bottom left panel shows the Eddington values computed with the FLD
approximation.  The bottom right panel shows $\rho$ for the same
two-dimensional slice. 

We first compare the SC and MC calculations which provide similar
results.  The consistency of the solution computed by these two very
different numerical methods strongly suggests that they are providing
accurate results. As can be seen for $f_{zz}$ in Figure \ref{f:fldmc}
the agreement between the radiation moments improves as the angular
resolution in the SC solver is increased (i.e. between the top left
and top right panels).  However, even with higher angular resolution
there are some modest discrepancies in the $f_{zz}$ near the
surface. This in part due to the statistical noise in the MC
calculation, for which S/N generally decreases as $z$ increases.  This
MC calculation was run with $\sim12$ billion photon packets with a
total computation time that exceeded the SC solver by a factor of
$\sim 100$. 

Since the improvement in S/N only increases as roughly $\sqrt{N}$,
where $N$ is the number of photon packets, further improving S/N
involves a substantial increase in the computational time.  Even for
this rather large number of photon packets, substantial noise remains
in the radiation field.  Such a high level of statistical noise could
lead to numerous problems when coupled to the MHD integrator.  Hence,
schemes which use MC methods to solve RT will generally require a
large number of packets.  Our results suggest that standard MC methods
need to be much more efficient or parallelized with effective load
balancing between the MHD integrator and the MC RT solver to be
competitive with SC methods when the simulation domain is far from
LTE\footnote{Although, there are problems where MC methods maybe
  preferable to SC, such as relativistic calculations that may
  require very high angular resolution if computed in the Eulerian
  frame.}.  Alternatively, it may be possible to significantly improve
on this performance by implementing some sort of hybrid MC scheme to
handle optically thick regions more efficiently
\citep[e.g.][]{Densmore:2007} since a significant fraction of the time
in our MC computation is spent solving RT in regions that are very
optically thick to scattering (so $f_{zz} \sim 1/3$) but still
optically thin to absorption.

There are several discrepancies between the SC/MC and FLD
calculations.  The most obvious is that with FLD, $f_{zz}$ approaches
unity by construction in the optically thin limit.  Obtaining $f_{\rm
  zz} = 1$, requires the radiation field to be concentrated in a
pencil beam of negligible solid angle around the $z$ axis, and is only
achieved on the $z$ axis at very large distances from a finite source.
Therefore, it is not appropriate for the upper boundary of a patch of
an accretion disk where the radiation field is still rather broadly
distributed over solid angle.  Indeed, $f_{zz} \sim 0.42$ is
consistent with estimates for a scattering dominated semi-infinite
atmosphere \citep{Chandrasekhar:1960}.  In principle, one could tailor
the flux-limiter to approach an alternative, problem dependent value,
although one can imagine applications where the appropriate limit will
be difficult to estimate a priori.

Furthermore, the FLD results yield $f_{zz} > 0.332$ everywhere, but in both
the MC and SC calculations $f_{zz} \lesssim 0.3$ is frequently
obtained in localized regions, consistent with a more horizontally
directed radiation field.  
It is also clear that the FLD Eddington factors correlate with $\rho$
to a much higher degree that in the SC or MC calculations.  Although
some correlation is present in the MC and SC calculations as well, it
is more prevalent in the optically thick regions and becomes much
weaker in the optically thin regions where the radiation field should
be more diffuse and more sensitive non-local variations in $T$ and
$\rho$.

Further discrepancies between the VET and FLD approaches are discussed
in JSD12.  The level at which these differences affect the overall
dynamics and thermodynamics remains unclear and ultimately requires
comparison with full numerical simulations using the SC/VET methods.
We note that the horizontally averaged flux in the SC and FLD methods
differs by $\lesssim 5\%$ at the top of the domain.  Hence the global
thermodynamic properties of the simulations may not be greatly
modified even though local properties of the radiation field differ.
Since simulations of accretion disk dynamics in the shearing box
approximation is one of our primary applications, we expect to be able
to make direct comparison with FLD-based results
\citep[e.g.][]{Hirose:2006} in the near future.

\begin{figure}
\includegraphics[width=0.47\textwidth]{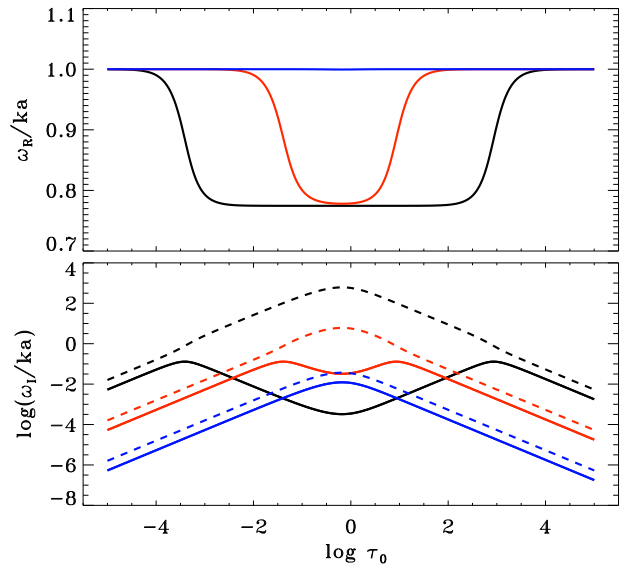}

\caption{
Real (top) and imaginary (bottom) parts of frequencies of radiatively
modified acoustic modes versus optical depth per wavelength $\tau_0$.
We plot three sets of curves and symbols corresponding to different
Boltzmann numbers Bo = 0.01 (black), 1 (red), 10, and 100 (blue).  The
curves correspond to radiation modified acoustic modes (solid) and a
non-equilibrium radiation diffusion mode (dashed).  We normalize
$\omega_R$ and $\omega_I$ by the product of the wave number $k$ and
the adiabatic sound speed $a$.  The real part of $\omega$ is zero
for the radiative diffusion mode.
\label{f:disp_ana}}
\end{figure}

\subsection{Radiating Linear Waves}
\label{linwave}

We now turn to tests of the RT solver when coupled to the MHD
integrator.  We first compute the radiative damping rate of linear
(acoustic) waves \citep{Stein:1967}.  The closely related problem of
the spatial damping of driven harmonic disturbances is covered in
\citet{Mihalas:1984}.  We briefly review the derivation of the dispersion
relation for such wave and refer the reader to these references for
further discussion.  We consider an ideal gas with a static, uniform
background state in LTE, with a grey absorption opacity $\chi$ and
frequency integrated thermal source function $B=\sigma_{\rm B}
T^4/\pi$.  Adopting the notation of \citet{Mihalas:1984}, we define
background and perturbed quantities with subscripts ``0'' and ``1''
respectively.  The background states has $\mathbf{v}=0$ with $\chi_0$
and $I_0=J_0=B_0$ constant everywhere.

With these assumptions and some algebra the linearly
perturbed versions of equations (\ref{eq:mass})-(\ref{eq:energy})
reduce to
\begin{equation}
\pdif{T_1}{t}-\left(\gamma-1\right) \frac{T_0}{\rho_0}\pdif{\rho_1}{t}
-\frac{4\pi (\gamma-1) \chi_0}{R \rho_0} \left(J_1 -B_1\right)=0,
\label{eq:t1}
\end{equation}
and
\begin{equation}
\left(\frac{\partial^2}{\partial t^2} - a^2_{\rm I} \nabla^2\right) \rho_1
-R \rho_0 \nabla^2 T_1 = 0,
\label{eq:rho1}
\end{equation}
where we $a_{\rm I}=a/\sqrt{\gamma}$ is the isothermal sound speed.  Similarly,
equation (\ref{eq:radtrans}) becomes
\begin{equation}
\hat{n} \cdot \nabla I_1 = \chi_0 \left(B_1- I_1\right).
\label{eq:linrt}
\end{equation}

\begin{figure}
\includegraphics[width=0.47\textwidth]{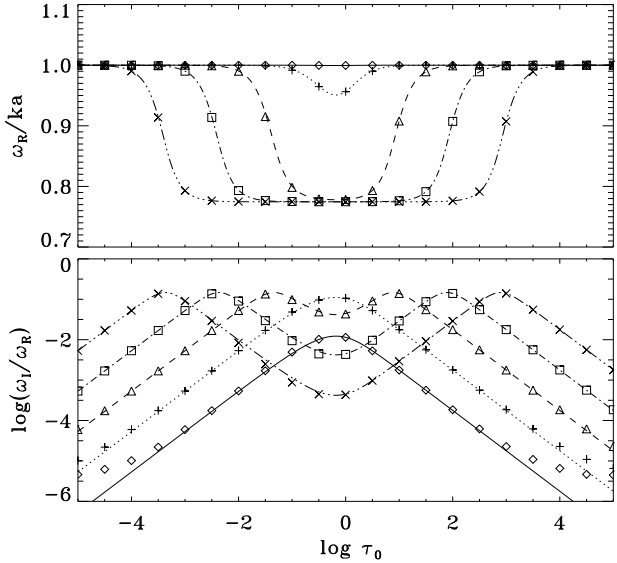}

\caption{
Real (top) and imaginary (bottom) parts of frequencies of radiatively
damped linear waves versus optical depth per wavelength $\tau_0$.  The
curves are analytic solutions to eq. (\ref{eq:disp}) and the symbols
are estimates derived from simulations. We plot four sets of curves
and symbols corresponding to different Boltzmann numbers Bo = 0.1
(dot-dashed, squares), 1 (dashed, triangles), 10 (dotted, crosses),
and 100 (solid, diamonds).  We normalize $\omega_R$ by the product of
the wave number $k$ and the adiabatic sound speed $a$.
\label{f:disp}}
\end{figure}

To linear order we can  assume
\begin{equation}
B_1  = \left(\pdif{B}{T}\right)_0 T_1 = 4 B_0 \frac{T_1}{T_0},
\end{equation}
and solve equation (\ref{eq:linrt}) directly to evaluate $J_1$ in equation
(\ref{eq:t1}).  We have
\begin{equation}
I_1 = \frac{4 B_0}{T_0} \int^{\infty}_0 T_1(\mathbf{x}_0 -\hat{n} s)e^{-\chi_0 s} ds, 
\end{equation}
where is $d s$ is a displacement parallel to $\mathbf{k}$.  We
consider plane wave solutions of the form $T_1 \propto e^{i (\omega t
  - \mathbf{k} \cdot \mathbf{x})}$.  Defining $\mu = \mathbf{k} \cdot
\hat{n}$ and integrating over solid angle, we obtain
\citep{Mihalas:1984}
\begin{equation}
J_1 =  \frac{4 B_0 T_1 }{T_0} \int^1_0  d\mu \int^{\infty}_0 dy 
\cos{(k \mu y/\chi_0)} \; e^{-y}.
\end{equation}
The integral evaluates to
\begin{equation}
J_1 = \frac{4 B_0 T_1}{T_0} \frac{\chi_0}{k} 
\tan^{-1}\left( \frac{k}{\chi_0} \right).
\label{eq:j1}
\end{equation}

We can now solve for the dispersion relation using equations
(\ref{eq:t1}), (\ref{eq:rho1}), and (\ref{eq:j1})
\begin{equation}
\omega^3- i \omega^2 \nu_0 \Xi_0 - \gamma a^2_{\rm I} k^2 +
i a^2_{\rm I} k^2 \nu_0 \Xi_0=0,
\label{eq:disp}
\end{equation}
in agreement with equation (16) of \citet{Stein:1967}.
We have defined 
\begin{equation}
\Xi = 1 - \frac{\chi}{k}  \cot^{-1}\left(\frac{\chi}{k}\right),
\end{equation}
and
\begin{equation}
\nu = \frac{16 \pi \chi B}{E_{\rm gas}},
\end{equation}
and the ``0'' subscript denotes that quantities are evaluated using the
background values.  To order unity $\nu_0$ is the reciprocal of the
radiative relaxation time in the background flow.

\begin{figure}
\includegraphics[width=0.47\textwidth]{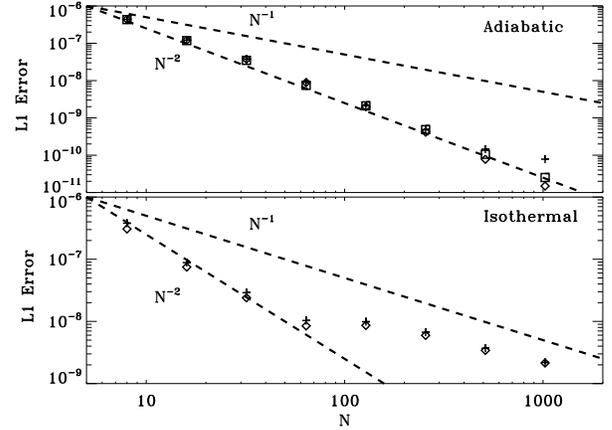}

\caption{
Convergence of the norm of L1 error as a function of resolution for
linear wave in a one-dimensional domain.  The top panel shows the
error norm for waves in the adiabatic regime corresponding to Bo=100,
$\tau_0=0.01$ (crosses); Bo=100, $\tau_0=100$ (squares); and Bo=0.1,
$\tau_0=10^4$ (diamonds).  The Bottom panels shows the error norm
for waves in the isothermal regime corresponding to Bo=0.01,
$\tau_0=0.01$ (crosses) and  Bo=1, $\tau_0=1$ (diamonds).  The dashed
curves show the expected trends for first-order ($N^{-1}$) and second
order ($N^{-2}$) convergence.  
\label{f:conv_1d}}
\end{figure}

Figure \ref{f:disp_ana} shows the solutions to equation (\ref{eq:disp}) for various
$\tau_0\equiv \chi_0/k$ (approximately the optical depth per
wavelength) and Boltzmann number 
\begin{equation}
{\rm Bo} =\frac{\rho_0 c_p T_0 a}{\sigma_{\rm B} T_0^4}=\frac{16 \gamma a \chi_0}{\nu_0}.
\end{equation}
Here $c_p$ is the specific heat at constant pressure, so Bo is the
ratio of the enthalpy flux (evaluated for $v=a$) to radiative flux.
There are two types of modes: radiatively damped acoustic waves (solid
and dashed curves) with phase velocity $v_{\rm ph} = \omega_R/k$
varying between $a_I$ and $a= \sqrt\gamma a_{\rm I}$ and a purely
damped ($\omega_R=0$) non-equilibrium radiation diffusion mode (dotted
curve).

The dimensionless ratio $\nu_0 \Xi_0/(k a_{\rm I})$ determines the
importance of radiation.  When this ratio is small equation
(\ref{eq:disp}) reduces to the standard adiabatic dispersion relation
with sound speed $a$ and the damping rate is approximately $\nu_0
(\gamma-1)/(2 \gamma)$.  For $\nu_0 \Xi_0/(k a_{\rm I}) \gtrsim 1$ the
phase speed $\omega/k$ decreases, approaching the $a_{\rm I}$ when $k
\sim \chi_0$ and $\nu_0 \gg a_{\rm I} k$, and the damping rate is
again small compared to $k a$.

For acoustic waves, the damping rate $\omega_I \lesssim a k$ for all
Bo and $\tau_0$.  However, this is not true for radiation diffusion
mode.  For $\tau_0 \sim 1$, $t^{-1}_{\rm rd} = \omega_I \sim \nu_0
\Xi_0$ and transitions to $t^{-1}_{\rm rd} \sim \nu_0 \Xi_0/\gamma$ for
$\tau_0 \gg 1$ or $\tau_0 \ll 1$.  Near $\tau_0\sim 1$, $\Xi_0 \sim 1$,
so the maximum decay rate has $t^{-1}_{\rm rd} \sim \nu_0$.  If $\delta t
> t_{\rm rd}$ then spurious, small-amplitude oscillations may grow
due to our failure to adequately resolve the radiative diffusion mode.

\begin{figure}
\includegraphics[width=0.47\textwidth]{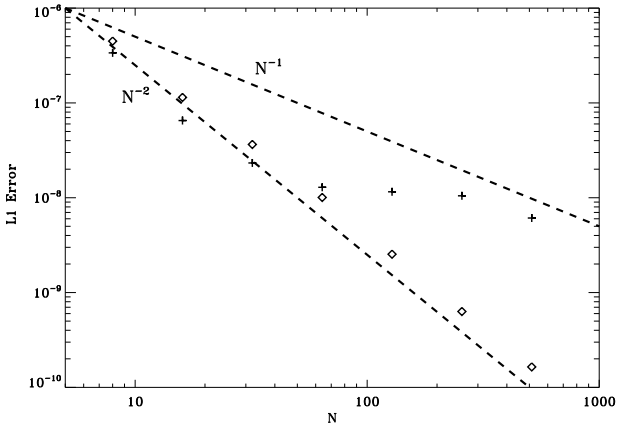}

\caption{ 
Convergence of the norm of L1 error as a function of resolution for
non-grid-aligned linear wave in a three-dimensional domain.  The crosses
represent waves in the isothermal regime (Bo=1, $\tau_0=1$) and the
diamonds in the adiabatic regime (Bo=100, $\tau_0=0.01$).  As
in the one-dimensional case, waves in the adiabatic regime converge at
nearly second order, but the isothermal waves are nearly second order
at low $N$, then plateau, and finally transition to a regime of
first-order convergence.  The waves are computed on a $2N \times N
\times N$ domain, as described in the text.
\label{f:conv_3d}}
\end{figure}

Indeed, we find exactly this type of numerical instability for a range
of $\tau_0$ if ${\rm Bo} \lesssim 1$.  The unstable range of $\tau_0$
corresponds to values for which $\delta t_{\rm C} \gtrsim t_{\rm rd}$
for modes with wavelengths comparable to the minimum grid spacing ($k \simeq
1/\Delta x$). Since we have an exact analytic solution for the
radiation source term from equation (\ref{eq:j1}), we can check this
result by turning off the RT solver and updating the total energy
using the exact expression for $Q_{\rm rad}$.  Even when the exact
expression is used the code is numerically unstable, as expected from
the argument above.  Limiting the time step to be less than or equal to
\begin{equation}
\delta t_{\rm rd} = {\rm min}\left(\frac{1}{\nu_i}
\left[1 - \frac{\chi_i \Delta x_i}{\pi} 
\cot^{-1}\left(\frac{\chi_i \Delta x_i}{\pi}\right)\right]^{-1} \right),
\end{equation}
stabilizes the solution when either the exact analytic expression or
the full numerical RT solution is used to compute $Q_{\rm rad}$.

Since $\nu \propto \chi$ this constraint is most stringent where
$\chi_i \Delta x_i \sim 1$ in which case  $\delta t_{\rm rd} = {\rm
  min}(1/\nu_i)$.  Assuming $\delta t_{\rm C} \simeq {\rm min}(\Delta x_i/a_i)$,
this implies that
\begin{equation}
\frac{\delta t_{\rm rd}}{\delta t_{\rm C}} \propto {\rm min}({\rm Bo}).
\end{equation}
Hence, whenever the Bo number in any gridzone of the domain is less than
unity, the maximum allowed time step will be determined by the radiation
constraint, unless some other physics (e.g. microphysical dissipation
or magnetic fields) enforces a shorter time scale.

We now use these solutions to evaluate the convergence properties of
the MHD integrator when our RT solver is used.  We simulate periodic
domains with different combinations of $\tau_0$ and Bo.  We initialize
the background with $\mathbf{v}=0$, $\rho_0=1$ and $\gamma=5/3$. The
initial perturbation is an eigenfunction with dimensionless amplitude
$A = 10^{-6}$.  We simulate
for one adiabatic crossing time $t_f =L/a$ and fit for the decay rate
and phase velocity.

Figure \ref{f:disp} shows a comparison of the numerically derived
dispersion relation with solutions of equation (\ref{eq:disp}).  Each
symbol corresponds to fits to a simulation of a one-dimensional domain
with $N = 256$.  Each curve corresponds to a different choice of Bo.
We find good agreement with theory for the phase velocities
$\omega_R/k$ and properly capture the transition from adiabatic to
isothermal and back to adiabatic as $\tau_0$ increases.  The agreement
for decay rates $(\sim \omega_I^{-1})$ is also good except for very
low or very high $\tau_0$ and high Bo.  In this case, the damping rate
$\omega_I^{-1}$ is very long compared to a wave period and higher resolution
is required to reduce the damping from numerical diffusion.

We now examine convergence properties in the characteristic regimes.
Figure \ref{f:conv_1d} shows the convergence of the norm of the L1
error vector, defined as
\begin{equation}
\delta q = \frac{1}{N}\sum_i |\mathbf{q}_i-\mathbf{q}^0_i|,
\end{equation}
where $\mathbf{q}^0_i$ is the eigenfunction used to initialize the
domain at $t=t_0$, but evaluated at $t=t_f$.  Each curve in Figure
\ref{f:conv_1d} corresponds to a set of simulations with different
combination of Bo and $\tau_0$.  The plotted simulations were run on
one-dimensional domains with $n_\mu=4$, but we obtain nearly
identical results for grid aligned waves in two-dimensional ($N \times
N$) and three dimensional ($N \times N \times N)$ domains.

Comparison with Figure \ref{f:disp} shows that all of the simulations
in the top panel are in the nearly adiabatic regime and those in the
bottom panel are in the nearly isothermal regime.  Since radiation has
only a small damping effect in the adiabatic regime, convergence is
nearly second order, as when radiation is entirely absent.  In the
isothermal regime, convergence is closer to second order at lower
resolution, but transitions to first order as resolution increases.
Since we use an operator split update of the energy equation, first
order convergence is expected when RT has a significant effect on the
thermodynamics.  Indeed, convergence is consistent with first order
when the time step is set solely by the CFL condition ($N \gtrsim
256$).  For $N \lesssim 128$, $\delta t=\delta t_{\rm rd} < \delta
t_{\rm C}$ and the radiation diffusion constraint sets the timestep.
In this case $\delta t$ is only very weakly dependent on $N$. 

We also considered the convergence of non-grid-aligned waves in two
and three dimensions.  The three-dimensional case is nearly identical
to the test presented in \citet{Gardiner:2008}.  We use a $2N \times N
\times N$ periodic domain, initialized with with a one-dimensional wave
that has been rotated with $\sin \alpha=2/3$ and $\sin \beta =
2/\sqrt{5}$ (see \citealt{Gardiner:2008}, for further details).  As in
the one-dimensional case, the initial wave is an eigenmode with
amplitude $A = 10^{-6}$ and we use $n_\mu =4$.  We again evolve the
domain for one adiabatic sound crossing time and evaluate the L1-error
norm via
\begin{equation}
\delta q = \frac{1}{2 N^3}\sum_{i,j,k} |\mathbf{q}_{i,j,k}-\mathbf{q}^0_{i,j,k}|.
\end{equation}

The convergence of the L1 error as a function of $N$ is shown for two
waves in Figure \ref{f:conv_3d}.  The solid and dotted curves show the
convergence for waves in the isothermal (Bo=1, $\tau_0=1$) and
adiabatic regimes (Bo=100, $\tau_0=0.01$), respectively.  Comparison
with Figure \ref{f:conv_1d}, shows that the convergence properties
are consistent with the one-dimensional/grid-aligned calculations.

Further linear wave tests are presented in JSD12, although these
assume the Eddington approximation and do not make use of the RT
solver employed here.  Since they solve the mixed frame moment
equations, the character of their numerically and analytically derived
dispersion relations differs from those presented here, although they
agree qualitatively in the appropriate limit.

\begin{figure*}
\begin{center}
\includegraphics[width=0.7\textwidth]{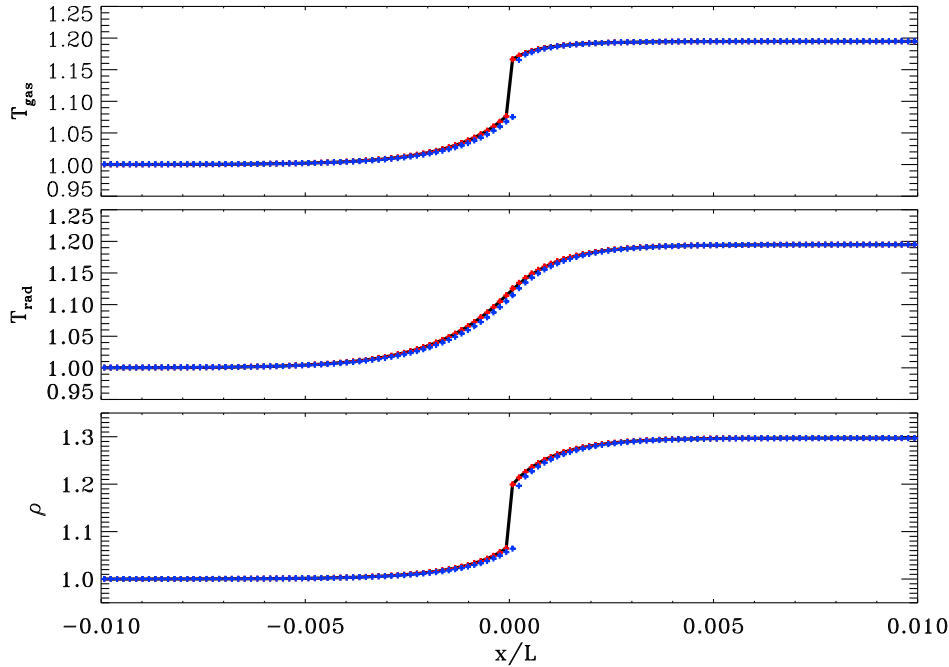}
\end{center}

\caption{ 
Profiles of the gas temperature (top), radiation temperature (middle),
and density (bottom) versus distance for a radiation modified shock
with $\mathcal{M}_0=1.2$.  The density, gas temperature, and velocity
are initialized with a semi-analytic planar shock solution computed
using the methods of \citet{Lowrie:2008} (shown as continuous curves
in each plot) while the initial radiation temperature is computed by
the RT solver.  Quantities are non-dimensionalized as described in
\citet{Lowrie:2008} and discussed in section \ref{radshock}.  The red
crosses indicate the initial conditions for all variables: the fact
that the radiation temperature computed by the RT solver initially
agrees with the semi-analytic solution confirms the accuracy of the RT
solver.  The blue crosses show the state of the variables after
evolving the shock for $t=L/a_0$.  There is a small drift in the
numerical solution due to the neglect of radiation pressure.
\label{f:m1.2}}
\end{figure*}

\subsection{Radiative Shocks}
\label{radshock}

We now consider the ability of the RT solver to model shocks in the
presence of radiation.  The physics of radiative shocks has been
explored by a number of authors \citep[see][and references
  therein]{Mihalas:1984} and is generally well understood.  However,
radiating shocks are sufficiently complicated that simple analytic
solutions for radiative shocks are generally not available.
Fortunately, \citet{Lowrie:2008} (hereafter LE08) have developed
fairly simple, semi-analytic methods for constructing one dimensional
planar solutions of radiating shocks, which are suitable for our
purposes.

LE08 construct their solutions using a grey non-equilibrium diffusion
model of radiation hydrodynamics.  Their treatment differs from ours
in a few important ways.  Rather than solving the RT equation
(\ref{eq:radtrans}) directly, they solve the radiation moment
equations with Eddington approximation and assuming a diffusion
relation for the radiative flux.  They retain a number of velocity
dependent terms which are absent in our treatment and include the
radiation source term in the material momentum equation (our
eq. \ref{eq:mom}). This allows them to explore the radiation pressure
dominated, which is not accessible with the methods discussed here
(see, however, JSD12).  Hence, our comparisons will be restricted to
shock solutions with a low ratio of radiation to gas pressure and
modest Mach numbers.

LE08 solve a non-dimensionalized systems of equations with solutions
that can be uniquely specified in terms of $\gamma$, $\sigma_a$, $
\mathcal{P}_0$, $\kappa$, and $\mathcal{M}_0$ using their
notation. Here $\sigma_a$ is the non-dimensional absorption cross
section, $\mathcal{P}_0$ is roughly the ratio of radiation to gas
thermal energy in the upstream flow, $\kappa$ is non-dimensional
photon diffusivity, and $\mathcal{M}_0=v/a_0$ is the upstream Mach
number.  The subscript ``0'' refers to upstream values in their
notation.

Following LE08, we examine solutions with $\gamma=5/3$,
$\sigma_a=10^6$, $\mathcal{P}_0=10^{-4}$, and $\kappa=1$.  In our
notation, these parameters correspond to $E_{\rm rad} = 10/9 \times
10^{-4} E_{\rm gas}$, $a = 1/\sqrt{3} \times 10^{-3} c$, and
$\chit=\chia=1/\sqrt{3} \times 10^{-3} L^{-1}$.  Here, $L$ is an
arbitrary reference length scale and all variables are
evaluated using their asymptotic upstream values.  

We construct one-dimensional planar shock solutions following the
procedures outlined in LE08 and use the resulting profiles of $\rho$,
$v$, and $E_{\rm gas}$ to initialize our one-dimensional simulation
domains.  Since $\chit L \ll 1$, we only simulate the region within a
few photon mean-free-paths ($\lambda_{\rm mfp} \sim 1/\chia \simeq
0.003 L$) of the shock front.  Since the semi-analytic
solutions rely on the Eddington approximation, we set $n_\mu = 2$
(i.e.  two-stream approximation) for consistency.  The radiation field
at the boundaries is fixed and assumes that the incoming radiation is in
thermodynamic equilibrium with appropriate upstream and downstream
asymptotic temperature.  We evolve the simulations for a time $\Delta
t = L/a$, which is typically a factor of hundred ($\sim L/\lambda_{\rm
  mfp}$) larger than the sound crossing time of the simulation domain.

\begin{figure*}
\begin{center}
\includegraphics[width=0.7\textwidth]{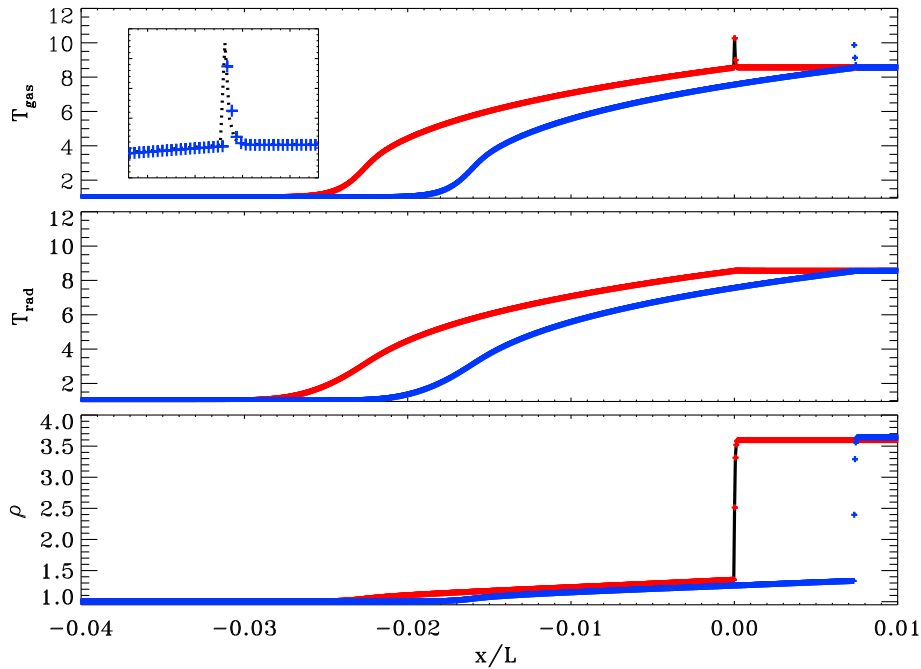}
\end{center}

\caption{ 
Same as Figure \ref{f:m1.2}, but for
$\mathcal{M}_0=5$.
The inset in the top
panel shows a close-up view of the gas temperature near the shock front at
$t=L/a_0$ as well as the initial semi-analytic model shifted to the right
by an amount $0.0073 L$ (the amount the shock front shifts in this timespan)
shown as a dotted line.  The shock shifts further since radiation pressure
is more important at this Mach number, nevertheless
the Zeldovich spike (and the overall shock profile) persists.
\label{f:m5}}
\end{figure*}

Figures \ref{f:m1.2} and \ref{f:m5} show characteristic results for
$\mathcal{M}_0=1.2$ and 5, respectively.  We use 128 and 1024 gridzones
for the simulation with $\mathcal{M}_0=1.2$ and 5, respectively.
We use a larger number for the $\mathcal{M}_0=5$ simulation to resolve
the narrow Zeldovich spike \citep{ZelDovich:1967}.  We plot the gas
temperature $T_{\rm gas} = a^2/(\gamma R)$, radiation temperature $T_{\rm
rad}=(E_{\rm rad}/a_R)^{1/4}$, and $\rho$ using the non-dimensional
units of LE08.  Here, $a_R$ is the standard radiation constant and
$E_{\rm rad}$ is computed using the RT solver.  The fact that $T_{\rm
rad}$ computed using the RT solver in the initial shock profile agrees
well with the semi-analytic solutions is already an important test of
our method.  Sufficiently far upstream or downstream of the shock,
the gas and radiation are in thermodynamic equilibrium with $T_{\rm
gas}=T_{\rm rad}$.  Near the shock front, the temperatures deviate,
with a radiation precursor upstream of the shock and Zeldovich spikes
appearing downstream of the shock for the higher Mach number solutions.

For each plot, we show two sets of curves corresponding to the initial and
final profiles.  Since we have initialized the simulation with stationary
solutions computed in the shock rest frame, the material properties
should not evolve with time.  However, since our system of equations
differ from those used by LE08 to derive their solutions (in particular,
we ignore the radiation pressure), our numerical solutions are only
approximately stationary.  The effects of the terms we have neglected
are small for the chosen parameters.  Nevertheless, there is a slow but
steady drift of the shock location in the downstream direction due to the
neglect of the radiation force in the upstream direction.  As the Mach
number of the flow increases, the radiation force becomes increasingly
important and the shock front moves more rapidly in this frame.
Even though the position of the shock drifts, the profile 
changes very little as the radiation source
term in the energy equation is still well approximated.

Further tests of radiative shocks are presented in sections 5.2 and
5.3 of JSD12, including calculations that use the RT solver to compute
the VET.

\subsection{Performance}
\label{perform}

The added computational cost of using the RT solver is determined by a
number of factors and will generally be problem dependent.  A useful
starting point is a comparison of the computational cost to integrate
the MHD equations for one timestep with the cost to perform a single
iteration of the RT solver for a single frequency when run on a single
processor.  For a three dimensional domain with $n_\mu = 4$ (i.e. 24
total rays) the RT solver requires $\sim 40$\% as many operations as
the CTU integrator.  This is essentially the simplest type of problem
that is of practical importance: an LTE grey problem with fixed
intensity on the boundaries and an angular discretization that can yield a
result beyond the Eddington approximation.

Many problems of interest will be more costly than this because we
will need multiple iterations, multiple frequency bins, or higher
angular resolution.  The total cost of the RT solution scales
approximately linearly with the number of frequency groups, total
number of angles, or number of iterations, all of which are problem
dependent.  Even for LTE problems, periodic boundaries or domain
decomposition may require multiple iterations.  For most problems
iteration will continue until the relative change in $S$ (or $J$) is
below some prescribed threshold $\delta_c$ and the total number of
iterations may fluctuate from one timestep to the next, depending on
conditions.

For the code tests considered here, which used $\delta_c = 10^{-5}$,
the number of iterations per timestep was $\le 5$, depending on the
problem, with 1-3 iterations being typical.  Most of the tests were
LTE and iteration was only used to handle boundary conditions.  An
exception is the uniform non-LTE atmosphere tests that were run with
$\gtrsim 1000$ iterations in order to obtain convergence of the
absolute error.

We emphasize that Figure \ref{f:conv_comp} is not indicative of the
typical number of iterations that need to be performed per timestep,
even in highly non-LTE domains.  The key point is that this
calculation starts from an initial condition that assumes an LTE
radiation field everywhere, even though the solution at the surface is
far from LTE.  As discussed in TF95, the main problem with the ALI
methods used here is that they have a rather small spectral radius.
Effectively, this means that it takes a rather large number of
iterations for errors in the initial condition that span many
gridzones to diminish.  Since we are computing RT on each timestep, we
already have an initial guess that is a reasonable approximation to
the correct non-LTE solution.  In particular, large (i.e. domain
scale) variations in the radiation field are usually already well
accounted for by the solution from the previous timestep.

We anticipate that our initial solution of the radiation field before
the first timestep may require hundreds to thousands of iterations for
highly non-LTE problems (i.e. those with a significant fraction of
zones having $\epsilon \ll 1$), but that subsequent timesteps will
only require a modest number ($\lesssim 5$) of iterations to obtain
relative convergence $|\Delta S|/S \lesssim 10^{-3}$.  Our initial
work on shearing box simulations (not reported here) supports this
expectation, although the number of iterations depends somewhat on
just how non-LTE the radiation field becomes.  \citet{Hayek:2010}
report similar numbers of iterations (see their Figure 2) as being
typical of their scattering dominated calculations.

A second consideration affecting performance is the maximum timestep
that can be used with the operator split update of the total energy
described in section \ref{interface}. The generalized CFL conditions
derived in section \ref{linwave} may reduce the timestep when $E_{\rm
  rad}$ is a significant fraction of $E_{\rm gas}$ or $v \ll c$.  For
such problems it will be more efficient to use the VET method of JSD12
when feasible.

A third consideration is scalability.  Since scaling efficiency will
be somewhat machine dependent, we are primarily concerned with
assessing how the code performs with RT relative to the (M)HD
configuration with no solution of RT.  To make the comparison
concrete, we study the weak scaling for $32^3$ gridzones per core on
the SciNet General Purpose
Cluster\footnote{http://www.scinet.utoronto.ca}, which consists of
eight core nodes made from two 2.53 GHz quad-core Intel Xeon 5500
Nehalem processors.  Tests were performed using the Infiniband
interconnect.  We use a grid aligned radiating linear wave on a
three-dimensional domain with $n_\mu=4$, $\tau_0=1$, and Bo=1 in the
radiating case and a slow magnetosonic wave for MHD only calculation.
Both tests were performed using the full MHD CTU integrator.  We
initialize the radiating wave as described in section \ref{linwave},
but we only allow a single iteration of the RT solver per timestep.
We compute the efficiency by dividing the number of zone cycles/second
obtained for a problem run with multiple cores by the number of zone
cycles/second for a single core.  The resulting scaling with number of
cores is nearly identical in the radiating and non-radiating cases,
falling to about 75\%\footnote{Note that this scaling efficiency for
  the non-radiating wave is somewhat lower than previous tests on
  other platforms.  See
  e.g. https://trac.princeton.edu/Athena/wiki/AthenaDocsScaling.}  at
512 cores (64 nodes).

We also find the same scaling efficiency for $n_\mu=12$, which is
notable because this correspond to a factor of seven increase in the
number of specific intensity bins that need to be passed on each
iteration.  In this case, the increase in communication demands
is balanced by the roughly factor of seven increase in the cost of
computing the RT solution with more angle bins.  These results suggest
that the scaling efficiency with the RT solver will tend to follow the
non-radiative scaling when they are run on the same platform, as long
as the number of iterations remains constant.  

The assumption that number of iterations stays fixed is an important
caveat.  In fact, this assumption will not generally hold since we use
iteration to handle subdomain boundaries.  To understand this, it is
useful to consider two LTE problems: one in which the optical depth
across each subdomain is very large and one for which the whole domain
is optically thin.  In first problem, the radiation field is
determined entirely locally and propagation of changes in the
radiation field from neighboring subdomains will require only a single
iteration.  In the second problem, variations in the emissivity on one
side of the domain can modify the radiation field on the other.  For a
cubic array of $N$ subdomains, it could take $\sim N^{1/3}$ iterations
to propagate the radiation across the entire domain.  This means that
scaling efficiency can, in principle, be very problem dependent.  For
most of the applications of primary interest, the majority of
subdomains will be optically thick so increasing the number of
subdomains should not significantly increase the number of iterations
required.  Therefore, scaling efficiency should reasonably consistent
with the non-radiating case.

Performance and scaling of the overall VET scheme is discussed in 
section 7.2 of JSD12.

\section{Summary}
\label{summary}

We have described our implementation of an RT solver in the Athena MHD
code.  Our module implements a short characteristic method for
computing RT on Cartesian, multidimensional simulation domains.  The
RT equation is solved once each simulation timestep for a
computational cost comparable to or less than a single timestep of the
MHD integrator for simple (e.g. LTE grey) problems.  Since we are
focused on astrophysical problems where velocities are slow compared
with the speed of light, we drop the time derivative of intensity and
the system becomes an integrodifferential equation with no explicit
time dependence for the radiation field.  The material properties of
the flow are evolved using the standard Athena MHD integrators, but
with radiative heating and cooling source terms computed from the RT
solver.  The code solves the RT equation for frequency dependent,
absorption and scattering opacities.  Non-LTE effects arising from
scattering processes are handled with ALI methods.  The resulting code
is well-suited for non-relativistic astrophysical problems where
diffuse emission, rather than bright point sources, constitutes the
dominant source of radiation.

We provide a detailed summary of the short characteristics and ALI
implementations. We also describe our implementation of an operator
split update of the energy equation using a radiation source term
computed directly by the RT solver.  Alternatively, the RT solver can
be used to compute a VET, which can then be input into the integration
of the coupled MHD and radiation moment equations.  The use of the RT
module for this purpose is discussed in a companion paper (JSD12).

We also present results from several test problems, which roughly fall
into two classes: tests of the RT solver on static simulation domains,
and tests of the coupled RT solver and MHD integrator for time-dependent
hydrodynamics simulations.  These tests demonstrate the accuracy of
the RT solver for multidimensional problems, assess its convergence
properties for applications where scattering leads to significant
deviations from LTE.  They indicate that substantial improvements in
accuracy and efficiency may be obtained over alternative methods, such
as flux-limited diffusion and Monte Carlo based RT solvers.

The tests also evaluate the accuracy and stability of the MHD
integrator when coupled to the RT solver via operator splitting.  They
verify that the code is generally only first-order accurate for
problems where heating or cooling of the fluid by the radiation field
is significant.  They also illuminate important time step constraints and
are useful for assessing the efficiency and accuracy of the code for
solving various astrophysical problems.

In particular, we derive a generalized CFL condition, predicated on
the need to resolve the non-equilibrium radiation diffusion mode.  The
requirement to place some limits on timesteps due to rapid radiative
relaxation \citep[see e.g.][]{Castor:2004} are generally acknowledged
and implemented in previous work.  However, we have not seen any
explicit reference to resolving the damping rate of the radiation
diffusion mode, which provides a a practical and precise guideline for
ensuring numerical stability.

The focus in this work has been on modeling the fluid dynamics and
thermodynamics with a self-consistent computation of the radiation
field.  However, we expect that the ability of the RT solver to
provide detailed outputs of the emergent radiation field, such as
images, lightcurves, and spectra, may be equally important.  Indeed,
for some applications the production of such outputs may be the
primary motivation for including radiation in the simulation. In
principle, the RT solver can be used {\it solely} to generate
diagnostic outputs, even in simulations without self-consistent
feedback of radiation on the material flow, either in real time or
via post-processing.

In addition to the simple test problems described here, we are
beginning a research program to simulate the local structure of
accretion flows (i.e. stratified shearing boxes) with radiative heating
and cooling.  Applications to radiation dominated environments using
the VET method (JSD12) are also underway, and include studies of
radiation dominated accretion disks, radiative Rayleigh-Taylor
instability, and the radiative driving of cold gas.  Applications to
extrasolar planets, star forming environments, boundary layers,
galactic and accretion disk outflow are also under consideration for
future work.

A significant limitation of the RT solver described here is the
short-characteristics method's inability to accurately handle bright
points sources.  We plan to address this in future work using a hybrid
scheme that computes the direct radiation from point sources using the
algorithms described in \citet{Krumholz:2007} and models the diffuse
emission with RT solver described here.

The source code for our RT solver, test problems and associated
documentation will be included in the publicly available version of
Athena\footnote{http://trac.princeton.edu/Athena} in the near
future. We have endeavored to make the code as user friendly as
possible and strongly encourage interested parties to use the code in
their own research.

\acknowledgements{We thank the anonymous referee for useful
  suggestions for improving the paper.  Ivan Hubeny provided considerate
  advice at the start of this project.  Bryan Johnson and Julian
  Krolik provided insightful questions and comments on an earlier
  draft of the paper.  We are also grateful to Shigenobu Hirose,
  Julian Krolik, and Omer Blaes for sharing their simulation results.
  Computations were performed on the GPC supercomputer at the SciNet
  HPC Consortium. SciNet is funded by: the Canada Foundation for
  Innovation under the auspices of Compute Canada; the Government of
  Ontario; Ontario Research Fund - Research Excellence; and the
  University of Toronto.  SWD is supported in part through NSERC of
  Canada.}

\end{document}